\documentclass[onecolumn, amsmath, amssymb, superscriptaddress]{revtex4}

\usepackage{dcolumn}% Align table columns on decimal point
\usepackage{bm}% bold math
\usepackage{amsmath}
\usepackage{amsfonts}
\usepackage{mathrsfs}  
\usepackage{graphics}
\usepackage[dvips]{graphicx}
\usepackage{times}
\usepackage{setspace}
\usepackage{color}   % {\color{red} TEXT }
\usepackage{verbatim}
\usepackage[raggedright]{titlesec}
\usepackage[normalem]{ulem}
\usepackage{framed}
\usepackage{enumerate}
\usepackage{footmisc}
\usepackage[breaklinks=true, hyperfootnotes=false]{hyperref}
\usepackage{breakurl}

\usepackage[graphicx]{realboxes}
\usepackage{varwidth}
\usepackage{balance}
\usepackage{flushend}
\usepackage{pdflscape}

\hypersetup{colorlinks=true,citecolor=blue, linkcolor=black,urlcolor=blue} % UNCOMMENT FOR SUBMISSION
\usepackage[outercaption]{sidecap} 

\graphicspath{{}}
\usepackage{hyperref}
\usepackage{amssymb}
\usepackage{mathbbol}
\usepackage{flushend}

\newcommand{\beginsupplement}{%          
	\setcounter{page}{1}
        \setcounter{section}{0} 
        \renewcommand{\thesection}{S\arabic{section}}%
        \setcounter{table}{0}
        \renewcommand{\thetable}{S\arabic{table}}%
        \setcounter{figure}{0}
        \renewcommand{\thefigure}{S\arabic{figure}}%
        \setcounter{equation}{0}
        \renewcommand{\theequation}{S\arabic{equation}}%
     }
     
\begin{document}
\title{Biomedical Convergence Facilitated by  the Emergence of  Technological and Informatic Capabilities}
\author{Dong Yang$^{a,}$}
\affiliation{Department of Management of Complex Systems, Ernest and Julio Gallo Management Program, School of Engineering, University of California, Merced, California 95343, USA}
\author{Ioannis Pavlidis}
\affiliation{Computational Physiology Laboratory, University of Houston, Houston, Texas 77204, USA}
\author{Alexander Michael Petersen$^{a,}$}
\affiliation{Department of Management of Complex Systems, Ernest and Julio Gallo Management Program, School of Engineering, University of California, Merced, California 95343, USA}

%\begin{abstract} 
%{\bf  \noindent  }
%\end{abstract}
% Abstract currently: 147 words
% NatureComm: 150 word abstract 
\maketitle

\footnotetext[1]{ %  \ \ $^{a}$  
\ \ \ These authors contributed equally; Send correspondence to: apetersen3@ucmerced.edu}

\vspace{-0.25in}
{\bf \small \noindent We analyzed Medical Subject Headings (MeSH) from 21.6 million research articles indexed by PubMed  to map this vast space of  entities and their relations, providing insights into the origins and future  of  biomedical convergence. Detailed analysis of MeSH co-occurrence networks identifies three robust knowledge clusters: the vast universe of microscopic biological entities and  structures; systems, disease and diagnostics; and emergent biological and social phenomena underlying the complex problems driving the health, behavioral and brain science frontiers. These domains  integrated from the 1990s onward by way of technological and informatic capabilities that  introduced highly controllable,  scalable and permutable research processes  and  invaluable imaging techniques for illuminating  fundamental   structure-function-behavior questions.  Article-level analysis confirms  a  positive relationship between team size and topical diversity, and shows  convergence to be increasing  in prominence but with recent saturation. Together,  our results invite additional policy  support for  cross-disciplinary team assembly to harness transdisciplinary convergence.}\\

\vspace{-0.1in}
% Knowledge networks
The codification  of  knowledge facilitates  more efficient search across the vast space of possible creative inputs accessible to scientists \cite{fleming2001recombinant} -- conceived in research as strategic  configurations of established and new entities, relationships,  tools,  equipment,  methods,   processes, observation, theory, etc.  Organization  of knowledge  into an ontology  facilitates additional understanding of its structure,  dynamics and  future trajectories   \cite{borner2010atlas,borner2015atlas,borner2021atlas}, as such knowledge maps \cite{borner2003visualizing,fleming2004science,borner2012design,shi2015weaving} improve scientists' ability to establish and recall complex relationships \cite{saket2015map}.  The pervasive drive to document entities and their relationships in detailed {\it *-omics} atlases will thereby help scholars manage the increasing volume and pace of  knowledge production \cite{pan2016memory,petersen_citationinflation_2018}  and accelerate breakthrough discovery \cite{rzhetsky2015choosing,helbing2012accelerating} by helping scholars  manage  the  uncertainty  associated with  exploration \cite{kuhn1959essential,march1991exploration}.  

% Biomedical innovation 
Against this backdrop,  we developed network-based methods for assessing the evolution of biomedical innovation.  
As the field of biology transitions from its descriptive roots into a  health innovation frontier, research increasingly combines tacit inputs  involving trial, error and practice (e.g. manual laboratory techniques)  with  explicit  inputs  that are highly transmissible and modifiable (e.g. pre-assembled computer algorithms). Consequently, the theory of recombinant innovation \cite{fleming2001recombinant} provides a powerful framework for understanding the potency, and uncertainty, embodied by   combinatorial approaches to search, refinement, experiment and discovery. 
Likewise, the triple-helix model of innovation \cite{leydesdorff1996emergence,etzkowitz2000dynamics,petersen2016triple} establishes the importance of catalysts  for bringing potent opportunities --  i.e.,  challenging problems  demanding novel solutions  met with a diverse supply of  research approaches -- to fruition. The catalysts in the present context are  techno-informatic capabilities, characterized by highly controllable,  scalable and permutable research processes for exploring and testing the exponentially vast number of  biological interactions. 

% CONVERGENCE as strategic cross-disciplinary configurations
In order to prime the pump, national innovation systems  have increased investment into transdisciplinary  {\it convergence science} \cite{NSFConvergenceAccel}, a paradigm prescribed by its originators as ``the coming together of insights and approaches from originally distinct fields'' \cite{NRC:2014,eyre2017convergence} -- as opposed to subfield integration characteristic of  interdisciplinary approaches \cite{colon2008chemical,pan2012evolution,Leahey_Sociological_2014}. Stimulation of potent triple-helix configurations aims to support the  emergence of new  hybrid disciplines \cite{sharp2011promoting},   for addressing  complex global challenges  \cite{helbing2013globally}. 
Recombinant innovation  is  fundamental to the {\it convergence science} value proposition, as integrating  diverse teams of experts   hedges against  uncertainty underlying the exploration process \cite{fleming2004perfecting} and provides a testable mechanism  \cite{Petersen:2018,HBP_2020} for explaining  the propensity of larger teams producing higher impact science  \cite{Wuchty:2007}. By way of example, the  advantage of  cross-disciplinary integration was evident in full force during the genomics revolution, wherein traditional biologists and computer scientists leveraged familiar operational language  to overcome relatively large  epistemic and cultural distances between their traditionally distinct fields \cite{Petersen:2018}.  
Recent work lends further support to   cross-disciplinarity \cite{fleming2004perfecting,Petersen:2018} as the more impactful  convergence  mode for exploring across long disciplinary distances \cite{HBP_2020}, as compared to expansive learning approaches \cite{Engestrom:2010}  pursued by mono-disciplinary teams.

% Main result 
% In what follows - what we do and what we find summary - highlighting Complementary Emergence of Techno-informatic Capabilities
It is generally appreciated that techno-informatic capabilities will transform many research domains, in addition to biomedicine. Yet even in the case of the latter, we lack  high-resolution maps for understanding the rich history  and future potential of  cross-domain integration. Hence, to identify and map the  anatomy of  biomedical convergence, we analyzed  millions of  intra-article keyword combinations  using  the high-resolution  Medical Subject Heading (MeSH) classification system \cite{MESH,MESHMap},  a controlled thesaurus  maintained by the U.S. National Library of Medicine (NLM) and implemented within the $\sim$30 million articles indexed by  PubMed.   
By visualizing MeSH co-occurrence networks and measuring inter-article MeSH diversity over the last half century, we identified the dominant knowledge configurations  underlying  contemporary convergence.
Notably, we document how  Technology (captured by MeSH branch J) and Informatics (MeSH Branch L)   emerged as convergence bridges,  transforming  non-invasive imaging, high-throughput  anomaly detection, and big data integration --  all key tools that are critical to addressing challenges at the frontiers of health,  behavioral and brain science \cite{HBP_2020}. 

%Figure 1 Here
\begin{figure*}
\centering{\includegraphics[width=0.9\textwidth]{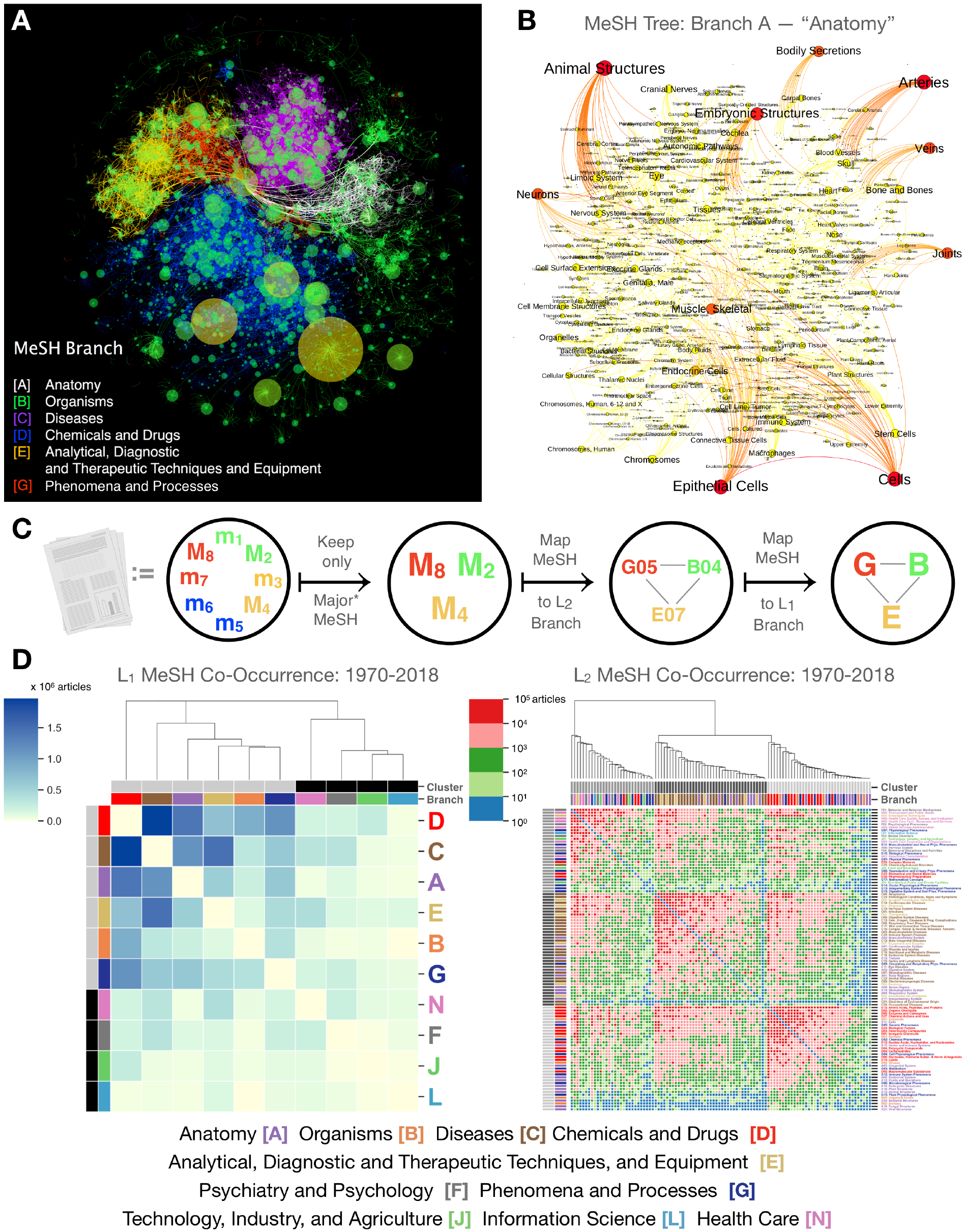}}
 \caption{  \label{Figure1.fig} {\bf Biomedical Knowledge Network.} 
 (A) Explicit network structure defined by the  Medical Subject Heading (MeSH) tree implemented by the US National Library of Medicine across the PubMed index. Visualized is the quasi-hierarchical MeSH-MeSH network composed from the six traditional biomedical branches. 
 (B) Network visualization of the ``Anatomy'' subtree (branch A). Nodes are individual MeSH and links are prescribed by the NLM MeSH Tree; nodes sized and shaded according to node degree. 
 (C) Serial refinement of the MeSH appearing within each PubMed article: the set of MeSH $\vec{m}_{p}$ for each article $p$ are refined to just the Major ``keywords'' (indicated within PubMed by an asterisk $*$); these Major MeSH $\vec{m}^{*}_{p}$ are readily mapped to their parent MeSH at the 2nd-level ($L_{2}$) and 1st-level ($L_{1}$).  (D) Historical MeSH co-occurrence frequencies for articles and reviews at the $L_{1}$ and $L_{2}$ levels; MeSH branches indicated by color-scale border segments (note the different color schema than in panel A). MeSH   clusters are determined by a modularity maximizing algorithm \cite{Blondel:2008} and indicated by the gray-scale border segments;  hierarchical structure indicates the minimum spanning tree representation of each co-occurrence matrix; entities are sorted within cluster in decreasing order of prevalence, calculated as the sum of co-occurrences with all other MeSH.}
\end{figure*}

\vspace{-0.2in}
\section*{Quantitative  Framework}
\vspace{-0.2in}
\noindent{\bf Related ontology and diversity research.}
Prior    scientific ontology studies use   journal and article classification systems of varying granularity and specificity levels. A common objective is to   measure disciplinary diversity by using classification systems as proxies for research topic subject areas, and measuring  category-category correlations \cite{wagner2011approaches,yegros2015does}. Some studies utilize broadly defined categories, such as the journal-level ``Subject Category'' descriptors implemented within Web of Science \cite{porter2007measuring,porter2009science,rotolo2013does}, or  article-level  systems used  within multi-disciplinary  journals    such  PNAS \cite{boyack2004mapping} and PLOS ONE \cite{petersen2019megajournal}. Another stream of research utilizes more  high-resolution article-level classification systems,  such as keywords  \cite{Leahey_Sociological_2014}, International Patent Classification (IPC) and US Patent Office Classification (USPC) codes \cite{fleming2001recombinant,fleming2007breakthroughs,youn2015invention,verhoeven2016measuring}, MeSH \cite{boyack2004mapping,boyack2011clustering,MESHMap,foster2015tradition,shi2015weaving,petersen2016triple} and Physics and Astronomy Classification Scheme (PACS) codes \cite{pan2012evolution}. Other approaches include  topic mapping based on co-word analysis  \cite{mane2004mapping} and hybrid approaches integrating keywords and cited references to group research into knowledge clusters \cite{borner2012design}.  \\

\noindent{\bf The MeSH tree.}  
We operationalize the representation of the biomedical knowledge network using  article-level MeSH annotations, which can be considered as keywords with varying specificity that are interrelated within a quasi-hierarchical tree  illustrated in  {\bf Fig. \ref{Figure1.fig}}(A).  
The official MeSH tree is maintained by the NLM and  consists of 16 branches designated by the characters A, B, C, D, E, F, G, H, I, J, K, L, M, N, V, Z (see   https://meshb.nlm.nih.gov/treeView to  explore the tree). The six branches  A, B, C, D, E, G represent core biological entities, concepts and methods; whereas the branches F, J, L, N represent  peripheral domains comprised also of entities,  concepts and methods. 

 Previous studies  focused on MeSH belonging to specific branch subsets, such as  C, D and E  \cite{MESHMap,petersen2016triple}. Here we exclude MeSH from the H, I, K, M, V, Z branches, as these are comprised of non-technical MeSH such as H (``Disciplines and Occupations'') and V (``Publication Characteristics'') -- see  {\bf Fig. \ref{Mtree.fig}} for a schematic representation of the quasi-hierarchical MeSH tree. To illustrate branch substructure,  {\bf Fig. \ref{Figure1.fig}}(B) shows the explicit MeSH-MeSH relations endowed in the  MeSH tree for branch A  (``Anatomy''). This particular branch is dominated by biophysical structures and their attributes -- i.e., representing one half of the structure-function dichotomy, with branch G (``Phenomena and Processes'') representing the other half. \\
 
\noindent{\bf Article-level Data.} 
 Our comprehensive analysis is based upon the  2020  PubMed index  consisting of more than 30 million index entries, with 93\% of these classified as ``Journal Article'' or ``Review''. We then pruned this dataset of articles lacking  Major MeSH  and focused on the sample spanning 1970-2018, resulting in  21.6 million publications (for additional details see {\it Methods}).  Regarding  notation, in what follows we use subscript $p$ to indicate article-level information such as publication year, indicated by $y_{p}$; the number of coauthors,  $k_{p}$; and the set of MeSH ``keywords'', represented as a vector $\vec{m}_{p}$ with 29,638 elements, each representing a distinct MeSH term. An article with  the sample average of 4 Major MeSH terms corresponds to $\sum_{i} \vec{m}_{p,i}=4$.\\

\noindent{\bf Projecting  MeSH onto 10 Subject Areas (SA).} 
We  analyzed all  MeSH belonging to the  10 core and peripheral branches. We refer to these  broad knowledge domains   as {\it Subject Areas} (SA).  These SA represent a  basis set for classifying  MeSH according to their  first-level parent branches (denoted by $L_{1}$). The MeSH tree is quasi-hierarchical (containing a relatively small number of loops), with 12 hierarchical levels  per branch \cite{MESHMap}. Hence, based upon this explicit relational tree structure,  we can  map any  MeSH occurring at the third level or greater to its corresponding second-level ($L_{2}$) MeSH term, for which there are 104 types. 

% #Mesh = 29638 and other #s: See Pubmed_ParseXML_Analyze.nb 
{\bf Figure \ref{Figure1.fig}}(C) illustrates the full process for producing $\vec{SA}_{p}$, which begins with first pruning out minor MeSH terms and auxiliary qualifiers, leaving just  Major MeSH terms that represent the article's core SA decomposition. By way of example, the   MeSH term  ``Obesity'' has  four Tree Number locators corresponding to three SA (C, E and G): C18.654.726.500; C23.888.144.699.500; E01.370.600.115.100.160.120.699.500; G07.100.100.160.120.699.500 (see  {\bf Fig. \ref{Mtree.fig}}). Hence, the $L_{2}$ representation of this single MeSH term is then given by \{C18, C23, E01, G07\} (which are  each distinct MeSH terms among the 29,638). We catalogue these MeSH using a $L_{2}$ count vector, denoted by  $\vec{SA}^{(2)}$, which contains 104 elements.  Further projecting this set to its  $L_{1}$ representation yields \{C, C, E, G\},  represented by  $\vec{SA}^{(1)}$=\{0,0,2,0,1,0,1,0,0,0\}. 
This example highlights the nuanced organizational structure of the MeSH tree, where terms can span multiple SA and contexts, thereby giving rise to relatively small yet non-trivial cross-branch connectivity within the explicit MeSH knowledge network -- see  {\bf Fig. \ref{MeshBranchABCDEG.fig}} for a network visualization of the explicit MeSH-MeSH relations within core biological SA. Combining multiple MeSH into $\vec{SA}_{p}$  yields a quantitative signature of each article's SA composition.

% Figure 2 Here
%\begin{landscape}
\begin{figure*}
\centering{\includegraphics[width=1.21\textwidth, angle = 90]{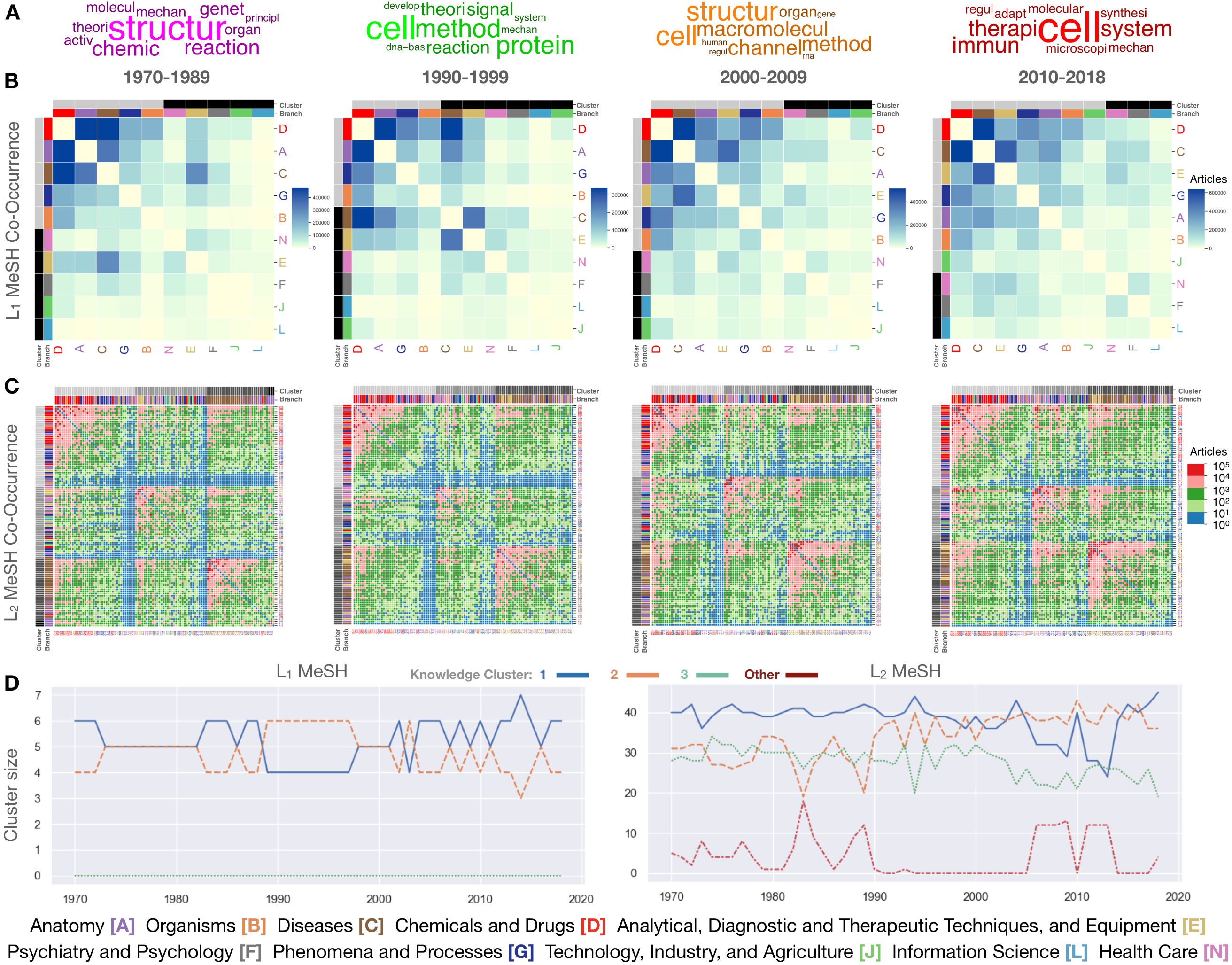}}
 \caption{  \label{Figure2.fig} {\bf Structure and Dynamics of  MeSH-MeSH Co-occurrence.}  (A) Top 10 most frequent destemmed keywords from Chemistry and Physiology or Medicine Nobel Prize rational statements, sized proportional to their prevalence, illustrating the transition from structural to mechanistic orientation of grand scientific pursuits. (B,C) MeSH branches are indicated by axis labels with corresponding color-scale border segments; MeSH clusters  determined by the Louvain modularity maximizing algorithm \cite{Blondel:2008} are indicated by the outer gray-scale border segments. MeSH are ordered within cluster in decreasing order of prominence corresponding to total number of article instances for a given period, as indicated by the numerical scales. (B) $L_{1}$ co-occurrence matrices show sequential  dynamics  across 4 non-overlapping periods; linear color-scale. (C) $L_{2}$ co-occurrence; logarithmic color-scale. (D)  Co-occurrence  cluster size dynamics  by year and resolution level: $L_{1}$ (left) and $L_{2}$ (right).
 }
\end{figure*} 
%\end{landscape}

\vspace{-0.2in}
\section*{Results}
\vspace{-0.2in}
\noindent{\bf  MeSH Co-Occurrence at the $L_{1}$ and $L_{2}$ levels.} 
 We assessed the entire space of   MeSH  combinations  by tallying SA co-occurrences among the present (non-zero) elements  contained in $\vec{SA}^{(1)}_{p}$ and $\vec{SA}^{(2)}_{p}$. Continuing with the example above, consider the SA counts  \{C, C, E, G\}, which could arise from an article with a single Major MeSH  annotation of ``Obesity'', or an article with four Major MeSH mapping individually to the same set of  SA; we do not distinguish between these two cases. We then tabulate the SA co-occurrences by counting for all unique SA-SA dyad types. In such a case, we tabulate three co-occurrence types: CE, CG and EG. When aggregating  co-occurrence tallies across articles, we combine normalized tallies -- e.g.  by assigning 1/3 weight to the CE, CG and EG matrix elements in the example above -- so that each article contributes a total weight of 1 to the co-occurrence matrix ${\bf M}^{(1)}$ (and similarly for ${\bf M}^{(2)}$ constructed at $L_{2}$). In this way, the total sum across  matrix elements of ${\bf M}$ is proportional to the total number of articles analyzed in the sample; see the {\it Methods} section {\it Major MeSH Descriptors}  for further details.

To begin, {\bf Fig. \ref{Figure1.fig}}(D) shows  co-occurrence frequencies recorded in the symmetric matrices ${\bf M}^{(1)}_{1970-2018}$  and ${\bf M}^{(2)}_{1970-2018}$, calculated across articles published between 1970-2018.  Each matrix indicates the SA name along the right border, accompanied by a consistent color indicating the corresponding $L_{1}$ branch, useful for guiding visual inspection. We grouped SA into knowledge clusters using the Louvain modularity maximizing algorithm \cite{Blondel:2008}, indicated by the gray-scale border segments along the upper and left matrix borders. The top border contains hierarchical clustering trees indicating the minimum spanning tree representation of each matrix. To further aid visual inspection, we maintain the  gray-scale cluster shades across  the aggregate matrices shown in {\bf Figs. \ref{Figure1.fig}}  as well as the time-disaggregated matrices visualized in  {\bf Fig.  \ref{Figure2.fig}}. 
  
As such, we analyzed MeSH co-occurrences at both the $L_{1}$ and $L_{2}$  levels, thereby producing knowledge network maps with complementary degrees of granularity. 
The main result for $L_{1}$  is the two-SA  substructure observed for ${\bf M}^{(1)}_{1970-2018}$. 
This substructure confirms that D, C, A, E, B and G form a core biomedical cluster, with the peripheral application domains N, F, J and L forming a second cluster. 

At higher resolution, the $L_{2}$  matrix ${\bf M}^{(2)}_{1970-2018}$ features three   mixed clusters. 
The first cluster is comprised of a wide array of $L_{2}$ MeSH pertaining to complex human phenomena related to behavior, physiology and health.
Importantly, this cluster also  includes L01 (``Information Science'') and J01 (``Technology, Industry, and Agriculture''). Because these are two focal domains in our analysis, we refer to their MeSH scope notes to provide additional context.
According to  MeSH scope notes, ``Technology''  refers to ``the science and application of techniques'' (J01), and more specifically ``the application of scientific knowledge to practical purposes in any field [including] methods, techniques, and instrumentation'' (J01.897). 
Likewise, MeSH scope notes describe ``Information Science'' (L01) as ``The field of knowledge, theory, and technology dealing with the collection of facts and figures, and the processes and methods involved in their manipulation, storage, dissemination, publication, and retrieval.'' 

The second cluster is  comprised of diagnostic methods associated with particular diseases and the systems they affect, as represented by MeSH primarily from  branches A (``Anatomy''), C (``Chemicals and Drugs'') and E (``Analytical, Diagnostic and Therapeutic Techniques, and Equipment''). 
The third cluster is comprised of biological entities and the phenomena that relate them, from cells to chemicals and the multi-scale  processes that connect  inputs, outputs and  characteristics of their reaction environments.\\

\noindent{\bf  Historical  co-occurrence trends.} 
The last half-century of biomedical research has witnessed incredible transition from a descriptive field -- focused around identifying  molecules, higher-order structures,  reaction pathways and mediators -- into a  field seeking to identify holistic mechanisms underlying systemic abnormalities in an effort to develop pointed therapies. This transition is illustrated through the Nobel Prizes awarded for \href{https://en.wikipedia.org/wiki/List_of_Nobel_laureates_in_Chemistry}{Chemistry} and \href{https://en.wikipedia.org/wiki/List_of_Nobel_laureates_in_Physiology_or_Medicine}{Physiology or Medicine}, which up until  the 1970s and 1980s predominantly  awarded  research  documenting key structural entities and reactions.  Based upon this early work, subsequent  Nobel research has transitioned  to 
addressing challenges at the nexus of the medical innovation triple helix --  where societal demand and industrial supply for acute solutions are facilitated  by  technological capabilities \cite{petersen2016triple}.

To illustrate this point,   {\bf Fig. \ref{Figure2.fig}}(A) shows the top ten stemmed keywords from Nobel Prize ``rationale'' statements, identifying prominent contextual themes across the last half century.
Indeed the 1980s brought forth  novel imaging, laboratory control and synthesis technologies, fundamental in the  1990s to map the genetic blueprints encoding  biological structure   \cite{Petersen:2018}. Polymerase chain reaction (PCR), among other  high-throughput  techniques,  accelerated science to the realm of  highly controllable,  scalable and permutable processes that  hitherto have been primarily restricted to traditional computational domains. Since then, the last twenty years has been marked by the rapid development of bioengineering capabilities, such as CRISPR gene editing tools \cite{doudna2014new} and powerful demonstration of stem cell pluripotency \cite{takahashi2007induction}, which allow scientists to more precisely understand  emergent structure-function relations, opening the door for  synthetic biology  applications \cite{benner2005synthetic}  that (re)design biological systems   \cite{church2014regenesis} or even develop altogether new  building blocks \cite{hoshika2019hachimoji}.   
% https://en.wikipedia.org/wiki/List_of_Nobel_laureates_in_Physiology_or_Medicine ; https://en.wikipedia.org/wiki/List_of_Nobel_laureates_in_Chemistry
%their regulation pathways

To visualize this history unfold, we  disaggregated the MeSH co-occurrence data  into  four non-overlapping periods: 1970-1989, 1990-1999, 2000-2009 and 2010-2018. For each period we  tabulate the co-occurrence matrix, denoted by ${\bf M}^{(1)}_{y}$ (${\bf M}^{(2)}_{y}$), where $y$ denotes a given period.
{\bf Figure \ref{Figure2.fig}}(B) illustrates the   structural dynamics at the $L_{1}$ level, identifying branch E and J as  vacillating domains switching between the two dominant clusters. The first $L_{1}$ cluster is characterized by  D, A, G and B -- largely descriptive SA, with the exception of ``Phenomena and Processes'' (G). The second cluster is comprised of N, F and L --  peripheral SA which are increasingly  convergent with the traditional SA, as indicated by   extreme off-diagonal elements representing cross-domain integration. 

At higher resolution, {\bf Fig. \ref{Figure2.fig}}(C) shows the structural evolution of ${\bf M}^{(2)}_{y}$, which is characterized by two relatively stable clusters and a more diverse and vacillating  intermediate cluster. Note that  MeSH are sorted within each cluster in decreasing order of prevalence, calculated as the sum of co-occurrence values across a given row of ${\bf M}^{(2)}_{y}$. Notably, ``Information Science'' (L01) and ``Technology, Industry, and Agriculture'' (J01) are located  in this intermediate cluster during the 1970-1989 period, where they play less dominant roles as indicated by their ranks within the cluster. However, over time their prominence within this cluster increases; by 2010-2018, J01  joined the biological agents and reactions  cluster, characterized by the core   ``Amino Acids, Peptides, and Proteins'' (D12),  ``Chemical Actions and Uses'' (D27),  ``Eukaryota'' (B01) and  ``Organic Chemicals'' (D02).  Close inspection reveals J01 having relatively strong co-occurrence with ``Inorganic Chemicals'' (D01), ``Investigative Techniques'' (E05), ``Health Care Quality, Access, and Evaluation'' (N06) and ``Biomedical and Dental Materials'' (D25) -- indicative of a highly convergent applications nexus facilitated by technological capabilities.

Close inspection further reveals a similar transformation for  L01, which  rose to prominence from the 1980s onward,  as genomics and other panoramic {\it omics} revolutions increased the demand for informatic solutions to map, characterize and associate entities into a functional atlas. This role did not merely manifest from increased technological capabilities associated with ``Diagnosis'' (E01) and ``Investigative Techniques'' (E05), but also involved the  integration of powerful ``Mathematical Concepts'' (G17), in particular ``Algorithms''  (G17.035) to  optimize classification and AI methods for ``Deep Learning'' (G17.485.500). Such methods are critical for capitalizing on new data-driven opportunities in Health Care  (represented broadly by N04, N05, N06) for  understanding   ``Behavior and Behavior Mechanisms'' (F01). \\

 % Figure 3 Here
\begin{SCfigure}
\includegraphics[width=0.52\textwidth]{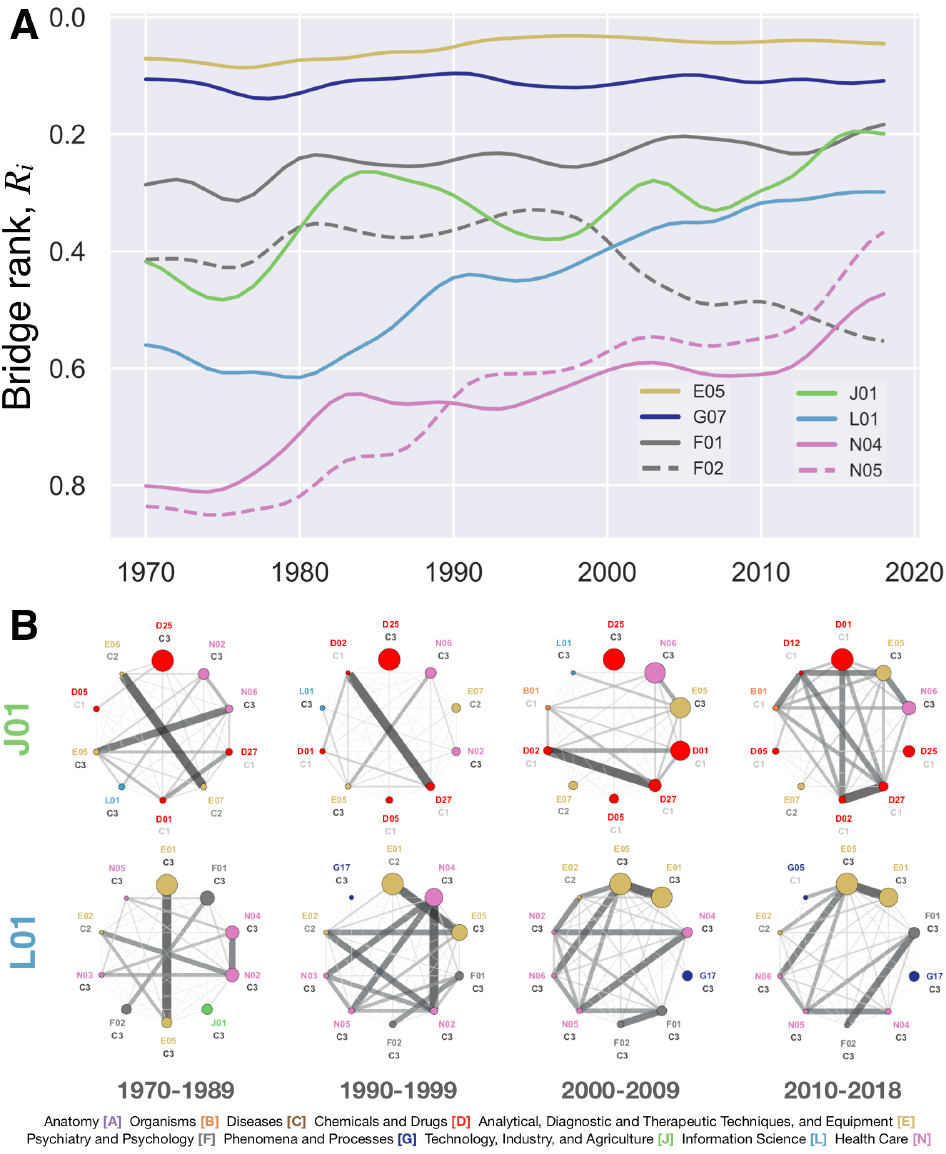}
\caption{  \label{Figure3.fig}    {\bf Prominent convergence bridges.} 
(A) For each MeSH, we analyzed the time series of bridge rank, $R_{i,t}$,  a normalized ranking based upon the knowledge bridge score ($\beta_{i}$) defined in Eq. (\ref{BS}). Smaller rank values ($R_{i,t}$) correspond to larger   $\beta_{i}$ values, representing MeSH that are highly co-occurrent with MeSH belonging to other knowledge clusters. 
Plotted are smoothed time series to facilitate visual inspection. 
Notably, J01 and L01 are rapidly emerging convergence bridges arising from the highly generalizable, scalable and codifiable nature of techno-informatic tools  and  algorithms facilitating novel non-invasive imaging, high-throughput analysis, measurement, and data integration. Other MeSH shown here identify the emerging convergence nexus of Health Care (N04, N05) and  Behavior (F01).
(B)  MeSH neighborhood subnetworks for J01 and L01 each show the ten most frequently co-occurring MeSH for the indicated period  -- sorted clockwise, starting from the top, with nodes sized proportional to  ${\bf M}^{(2)}_{ij}$; each MeSH's SA  is indicated by  its node/label color.
Links are plotted with thickness and shade proportional to ${\bf M}^{(2)}_{jj'}$, thereby indicating the cross-domain linkages among prominent neighbors that are facilitated by $i$.   Each node includes its MeSH identifier and a knowledge cluster identifier, the latter indicated by a gray  scale gradient. 
For example, MeSH J01 ``Technology, Industry, and Agriculture'' (which is a member of C3 for the first three periods and subsequently transitioning to C1 in the most recent period) is highly connected to MeSH from all other  clusters (C1-C3), in particular to L01 until its disassociation in the most recent period 2010-2018; interestingly, L01 ``Information Science'' diverged from J01 as early as the first period 1970-1989, subsequently becoming more strongly coupled with members of branch E and G. See {\bf Figs. \ref{FigureSIBranchScore.fig}-\ref{FigureSINNetworks.fig}} for extended analysis of bridge ranks and neighborhood subnetworks for the remaining convergence bridges
}
\end{SCfigure}

%{\color{magenta}
\noindent{\bf Biomedical convergence identified by the emergence of cross-cluster  bridges.} 
 To identify MeSH  serving as bridges that link different knowledge clusters, we developed a metric which quantifies the degree to which individual nodes  connect distinct clusters in a connected weighted network. This metric is motivated by the bridge centrality index \cite{jensen2016detecting}. We develop this approach with ${\bf M}^{(2)}_{t}$ in mind, which is a  complete network as opposed to a sparse network (See  {\it Methods} section {\it Calculation of the cross-cluster bridge score}  for further details). Applying this method to ${\bf M}^{(2)}_{t}$ yields a knowledge bridge score for each MeSH and year, denoted by $\beta_{i,t}$, which is large if a MeSH is highly connected to clusters other than its own.  
 
 {\bf Figure \ref{Figure3.fig}} shows the rank dynamics $R_{i,t}$ (where larger $\beta_{i,t}$ correspond to smaller rank $R_{i}$) for 8 MeSH  that are  distinguished by either persistent growth or sustained prominence  as knowledge bridges over the entire study period. Notably, L01  features steady growth following its emergence in the early  1980s, whereas   J01 features an oscillating upward trend. We also observe a divergence  between  ``Behavior and Behavior Mechanisms'' (F01) and ``Psychological Phenomena'' (F02), with the latter becoming increasingly insular over the last 20 years.
 The emergence of  health care bridges  relating to  ``Health Services Administration'' (N04), ``Health Care Quality, Access, and Evaluation'' (N05) and situational behavior analysis (F01) is another disruptive trend from the last decade -- consistent with the ramp-up of several  international {\it Human Brain Projects}  \cite{Markram:2012,Grillner:2016}  in 2013,  thereby championing the   brain research nexus as the next  frontier \cite{HBP_2020}.\\

% Figure 4 Here
\begin{figure}
\centering{\includegraphics[width=0.85\textwidth]{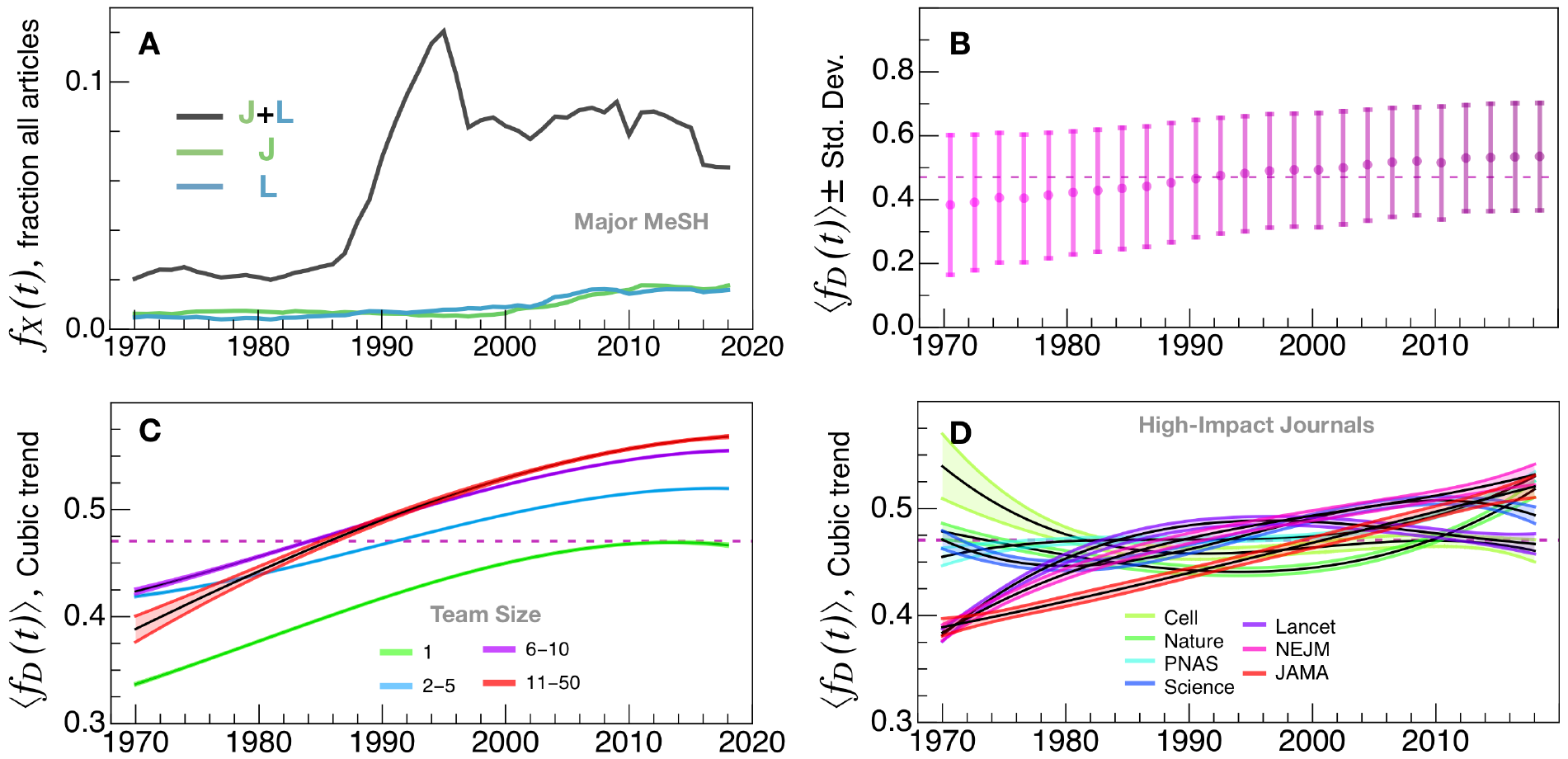}}
\caption{ \label{Figure4.fig} {\bf  Increasing prevalence of  cross-domain  knowledge integration.}   
(A) Fraction  of articles, $f_{X}(t)$,   featuring at least half of Major MeSH in ABCDEG and at least a quarter in  J, L or J+L. 
(B) $f_{D,p}$ measures  article-level SA diversity,  based on weighted pairwise  SA-SA combinations defined in Eq. (\ref{fDp}). Average $f_{D,p}$ diversity value and standard deviation (error bar) calculated    for non-overlapping 3-year windows over the period 1970--2018; the global  average  diversity value (0.47) is indicated by the horizontal dashed line. 
The mean value $\langle f_{D}(t) \rangle$ has increased from 0.38 to 0.53, representing 39.5\% growth, over the entire 49-year sample period.
(C)  Increased knowledge diversity correlates with the emergence of team science, where the most rapid increase is among the larger teams with 11-50 coauthors. 
Solo-author research has only recently  reached the  average diversity value indicated by the horizontal dashed line.
(D) Heterogenous   trends   by  longstanding core (Cell), elite multi-disciplinary (Nature, PNAS, Science)  and biomedical (Lancet, NEJM, JAMA) journals. NEJM and  JAMA show the largest  consistent increase over time; Nature and PNAS exhibit the largest increase  in the last decade. %; Science, Lancet and Cell  are stable.
%, with the latter featuring overall decline across the entire period, consistent with the journal being the most mono-disciplinary of the representative group. 
Each  curve represents the third-order polynomial trend, with shaded colored areas indicating the 99\% CI for each trend line.}
\end{figure}

\noindent {\bf  Convergence factors  -- SA composition, prevalence, team size and scientific impact.} 
 The value proposition envisioned by convergence science originators \cite{NRC:2014} was novel configurations of experts strategically assembled  to address a dimension of the underlying problem that would be otherwise inaccessible by mono-disciplinary strategies.  The convergent union between biology and computing experts in the genomics revolution provides a rich example \cite{Petersen:2018}. Hence, it follows that convergent research strategies are more likely to prevail in  team science endeavors, and provides a testable mechanism for the propensity for larger teams to produce higher-impact science  \cite{Wuchty:2007}.

Indeed, the importance of informatic (L) and technological  (J) capabilities to modern biomedical research cannot be understated. To illustrate the emergence of these bio-technological and bio-informatic  modes,  indicated  by $X$, we calculated the fraction of articles by year that contain a significant component belonging to SA J and/or L. To ensure that the articles are  otherwise focused on traditional biomedical SA, we estimate  the number of cross-domain articles by focusing on the subset of PubMed articles featuring a half or more of their MeSH in the core categories ABCDEG. We then assign the indicator $X$ to those articles which also contain at least a quarter of their MeSH belonging to  L, J, or L+J in combination, and report the quantity of articles featuring $X$ as the fraction  $f_{X}(t)$ of the total articles by year. 

{\bf Figure \ref{Figure4.fig}}(A) indicates a burst of activity around the early 1990s  for research  containing  both L and J in combination -- much greater than the $f_{X}(t)$  calculated for either SA alone. This particular configuration of ABCDEG $\times$ J $\times$ L represents a formidable nexus featuring the combination of  high-throughput equipment to produce and churn through massive biomedical data.   For comparison,  {\bf Fig. \ref{FigureSIfD.fig}}(A) applies the same method to all MeSH (i.e., not distinguishing between Major and Minor MeSH), thereby including  the more  peripheral MeSH capturing the idiosyncratic research details. This second perspective indicates a more steady integration over time of J and L capabilities at the SA periphery, with the strongest upturn in J+L in the early 2000s, which has since saturated. Hence, if the  1990s brought forth  the revolution in biomedical research, and the 2000s witnessed the further penetration of this paradigm shift, there appears to be a recent stagnation owing primarily to L which has permeated other convergence zones \cite{Roco:2013}, such as computational social science \cite{lazer_computational_2009} and the science of science \cite{Fortunato_2017_Science}. 

 To further explore the evolution of cross-domain SA configurations at the article level, we developed a Blau-like  diversity measure, denoted by $f_{D,p}$, and  based upon SA-SA co-occurrence. This  measure  accounts for  two distinct  diversity types  -- both categorical  variation and  concentration disparity \cite{harrison2007s}. 
Our method takes as input the categorical count  vector $\vec{SA}_{p}$ and applies the outer tensor product, $\vec{SA}_{p} \otimes \vec{SA}_{p}$,  yielding a weighted matrix ${\bf D}_{p}$ that captures dyadic SA-SA co-occurrence;   see {\it SI Appendix}  Eqs. (\ref{Dmatrix})-(\ref{fDp}) for further elaboration. To summarize, if an article's MeSH descriptors are  contained in just a single SA, then there will be  only one single non-zero value contained in ${\bf D}_{p}$, which  will occur along the matrix diagonal. Conversely, if the article features several SA, then   off-diagonal elements represent cross-domain combinations.  Hence, $f_{D,p}$ measures the relative fraction of off-diagonal  to diagonal elements  -- with $f_{D,p}=0$ corresponding to minimal diversity.  
 Because more evenly distributed SA counts  will yield relatively  larger off-diagonal values, and hence larger $f_{D,p}$ value, it is both a measure of   variation and disparity  \cite{harrison2007s}. Morever, $f_{D,p}\in [0,1)$ is a  bounded statistic with  clear interpretation of the  lower and upper bounds.

We computed $f_{D,p}$ based upon $\vec{SA}^{(2)}_{p}$ counts tallied for each publications in our PubMed sample (i.e. using the $L_{2}$ MeSH representation of an article). {\bf Figure \ref{Figure4.fig}}(B) shows the evolution of the mean diversity value $\langle f_{D}(t) \rangle $ calculated for  articles grouped by publication year, and shows a  steady increase in  SA diversity over the last half  century. Analysis of the entire distribution of values, $P(f_{D,p,t})$, also exhibits a systematic positive shift across the range of  $f_{D,p}$ values, and so the increase in the mean value is not just the result of an uper tail effect in the aggregate distribution. Rather, we observe a marked decrease in the prevalence of mono-SA articles characterized by $f_{D,p}=0$ in the lower distribution tail -- see  {\bf Fig. \ref{FigureSIfD.fig}}(B). This result is analog to the reduced frequency of single-authored research observed  over the same period \cite{Wuchty:2007}, indicative of  the  intense burden to integrate distant knowledge domains, a challenge that contributes to the disappearance of the ``renaissance''  solo genius  \cite{Heller:2015,simonton_after_2013}. 

 To assess how team size mediates article-level SA diversity, we  disaggregated the data into four team-size groups: solo-author (coauthor number $a_{p}=1$); small-sized team  ($1 < a_{p}\leq 5$); medium team  ($5 < a_{p}\leq 10$); and large team ($10 < a_{p}\leq 50$). {\bf Figure \ref{Figure4.fig}}(C) shows the dominant trends in average diversity time series $\langle f_{D}(t)\rangle$ for each team size group, from 1970 to present. To facilitate visual comparison of the dominant trends, shown are each time series fit using a third-order polynomial: $\langle f_{D}(t)\rangle \equiv a + b(t-1990) + c(t-1990)^{2}+d(t-1990)^{3}$. Notably, the curve for solo-author team shows a significant systematic shift towards  lower SA diversity values. From the mid-1980s onward, the curves are ordered according to team size group, with the largest teams achieving SA diversity  levels far in excess of the unconditional average for the entire period, $\langle f_{D} \rangle = 0.47$ (indicated by the horizontal dashed line in each figure sub-panel). As a robustness check, {\bf Fig. \ref{FigureSIfD.fig}}(C) shows the journal-level relation between article-level SA diversity and mean journal team size for the  60 biomedical journals appearing in the set of top-100  journals ranked by  2018 Clarivate Analytics JCR Impact Factor, indicating a higher range of SA diversity associated with larger teams.
 
The wide variation in journal-level $\langle f_{D} \rangle_{j}$  indicates that support for convergence science is a feature of the distinct communities of expertise. While it is beyond the present study, we speculate that the multidisciplinary composition of a journal's editorial board  is likely to be a strong factor underlying the prevalence of convergence science featured in a particular journal.  Among the most prestigious journals shown in  {\bf Fig. \ref{Figure4.fig}}(D), several feature recent decline in  $\langle f_{D}(t)\rangle $ over the last two decades, including  Cell and the Lancet; contrariwise,   the medical journals  NEJM and JAMA  are consistently trending upwards. To further illustrate the dynamics at the journal level, {\bf Fig. \ref{FigureSIfD.fig}}(D) shows the top-10 journals in terms of their largest growth in $\langle f_{D}(t)\rangle $ over the study period, featuring prominent journals in multiple domain areas including cancer, microbiology, medicine and psychology.  Despite such consistent growth trends for some journals, this is not a universal feature. See {\bf Table \ref{table:top20}} for prominent journals ranked according to  their average SA diversity value $\langle f_{D} \rangle_{j,1970-2018}$.

\vspace{-0.2in}
\section*{Discussion}
 \vspace{-0.2in}
 %Preamble about knowledge networks to organize entities and concepts, entities in relational heirarchies 
With each new   discovery -- some large, others incremental -- there is also a need to  find its place in the order of things --  some   adding new layers to our understanding, and others filling in gaps.
Classification systems are designed to  manage the increasing volume of knowledge, so   that it can be readily recorded, searched, explored and exploited in the pursuit to create new knowledge. 
From the Decimal Classification system (developed by expert committees and used in libraries around the world)  to the Wikipedia category structure (a  crowdsourced representation of our collective knowledge  \cite{UDCWiki1,UDCWiki2,UDCWiki3}), we are surrounded by ontologies that attempt to organize our common understanding. Likewise, the MeSH system aims to organize biomedical knowledge into a networked  hierarchy that relates objects, methods, theory and other contextual metadata; notably, the traditional PACS system  has recently been revamped into  PhySH by adopting a similar concept-based hierarchy that is better suited for integrating the multiple sub-disciplines and facets that define the domain of physics   \cite{PhySH2019}.

% Convergence
We exploited the MeSH ontology structure to develop a dynamic and high-resolution  map of cross-domain integration, which contributes to prior efforts to visualize the biomedical  knowledge network \cite{MESHMap,petersen2016triple,shi2015weaving}.
% Key Results: Multi-level analysis of knowledge network  and article level diversity 
In particular, we developed methods to visualize   MeSH co-occurrence  and quantify cross-domain diversity.
Macro-level analysis  of MeSH co-occurrence networks identifies three robust knowledge clusters: the vast universe of microscopic biological entities and  structures; systems, disease and diagnostics; and  biological and social phenomena capturing emergent properties, processes and functions. Research at the health, behavioral and brain science frontiers typically integrate these knowledge domains, 
 signaling the  emergence and future potential of convergence science \cite{NRC:2005,sharp2011promoting,Roco:2013,NRC:2014,ConvergenceElephant_2014,eyre2017convergence,HBP_2020}. 
The convergence nexus of Health Care (N), Behavior and Behavior Mechanisms (F01) and Information Science (L01) is a prime example, making way for transdisciplinary brain science  \cite{HBP_2020} to map and model brain circuits \cite{Jorgenson:2015} that are fundamental to addressing the grand challenge regarding the  ``global burden of mental disorders'' \cite{Kessler:2009,dzau2018reimagining}. 
We reinforced these macro trends   by micro-level analysis of article-level MeSH diversity,  which  increased by 40\% over the entire study period from 1970 to 2018 -- see {\bf Fig. \ref{Figure4.fig}}(B).
This shift can  be partly explained by a reduction in the prevalence of mono-domain research, analog to the simultaneous reduction of solo-authored research \cite{Wuchty:2007}.

% Key Result:  Comprehensive analysis of Knowledge Network Identifies periods of disruption:
Comprehensive analysis of the knowledge network can also identify periods of  disruption.
By assessing the dynamics of  MeSH usage across the entire PubMed index, we identified  periods in which MeSH increasingly fluctuate between knowledge clusters  (see {\bf Fig. \ref{FigureSIStableCliques.fig}}). 
To identify the particular  MeSH that bridge different knowledge communities we  developed a cross-cluster bridge metric in Eq. (\ref{BS}).
% Key Result: Fig 3 - emergence of J01/L01 as convergence bridges
{\bf Figure \ref{Figure3.fig}} shows the emergence of technology (J01) and  informatics (L01) as prominent convergence bridges, reflecting the highly generalizable, scalable and codifiable  characteristic of techno-informatic tools  and  algorithms that accelerate scientific discovery by facilitating  high-throughput, measurement, data integration and analysis. 
% Key Result: Fig 4A,B
{\bf Figure \ref{Figure4.fig}}(A) further indicates the prevalence of this potent combination, which rapidly emerged in the early  1990s during the genomics revolution. 
Notably, this endeavor highlights the role of  cross-disciplinary collaboration between scholars from traditional biology and computer science departments, which provided an early example  of successful convergence,  likely owing to the consortium science model whereby teams of teams  organize  with a common goal to share benefits equitably within and beyond institutional boundaries   \cite{Petersen:2018}.
% Key Result: Fig 4 - Team perspectives
In related work, we show that research integrating  subject areas that span relatively  larger disciplinary distances is more impactful when executed by  cross-disciplinary teams as opposed to mono-disciplinary teams \cite{HBP_2020}.
We provide complementary support, showing that medium and large-scale teams have a notable advantage  integrating multiple research domains -- see {\bf Fig. \ref{Figure4.fig}}(C).
Viewed from a different bureaucratic perspective, we observe  variable levels and growth rates of  convergence in journals, possibly owing to the multi-disciplinary composition (or lack thereof)  of journal editorial boards.

In summary,   techno-informatic capabilities  are increasingly essential to the biomedical frontier. 
As illustrated by {\bf Fig. \ref{Figure4.fig}}(A), these distinct  inputs are increasingly found as complements rather than substitutes -- for the period 2000-2018 we estimate that  $\langle f_{\times} \rangle \approx $8.2\% of research articles  relied on techno-informatic capabilities in tandem (J+L); this frequency corresponds to a 291\% growth over the J+L convergence levels in 1980.
If the future relies on harnessing the combined power of human and machine intelligence, then the development of {\it human-in-the-loop} and Man-Machine Systems (J01.897.441) ``in which the functions of the man and the machine are interrelated and necessary for the operation of the system'' will require  transdisciplinary domains such as bio-mechatronics to flourish  by harnessing convergence \cite{li2016guest}. For this reason, scientists are increasingly in need of knowledge maps to navigate the realm of possibilities and to thereby  which conceptual bridges to cross and which to  avoid.

{\small
 \vspace{-0.2in}
\section*{Methods}
 \vspace{-0.2in}

\noindent{\bf Subject Area (SA) classification using MeSH.} 
We used the 2020 MeSH classification tree, which includes 29,638 descriptors   that are organized in a tree-like structure denoting their hierarchical relations. MeSH descriptors are assigned to biomedical publications through an indexing process performed by examiners at the US National Library of Medicine (NLM).
 With on average 12 MeSH per article, this ontology facilitates topic mapping and topic co-occurrence analysis at multiple levels of specificity \cite{petersen2016triple}. We restrict our analysis to  the  ``Major Topic Headings'' for each article, which are indicated in each PubMed article page by an asterisk  $^{*}$ next to the MeSH term; these Major MeSH  account for roughly 1 in 3 MeSH descriptors, and so with on average 4 Major MeSH these annotations  are   sufficient to identify the article's core content. As such, we use these publication-level MeSH  to determine the topical Subject Areas (SAs), as indicated by the 10 pre-defined science-oriented MeSH branches (A, B, C, D, E, F, G, J, L, N).\\ 

\noindent{\bf Major MeSH Descriptors: refinement to 1st and 2nd branch level representation.} 
For each publication $p$, we extracted the set of Major MeSH terms  (as indicated by an asterisk in  PubMed records) and represent them by the vector $\vec{m}_{p}$ containing 29,638 elements, each representing a distinct MeSH term. We then map each MeSH term  to its corresponding tree location, which  specifies its classification among the 10  science-oriented MeSH  branches. To be specific, we define a generic operator $O_{M}$ which takes the vector $\vec{m}_{p}$ and maps these counts to a combined count vector $\vec{SA}^{(2)}_{p}$ with   104 elements representing all  $L_{2}$ MeSH:  $O^{(2)}_{M} (\vec{m}_{p}) = \vec{SA}^{(2)}_{p}$. Similarly, we define the operator $O_{M}^{(1)} (\vec{SA}^{(2)}_{p}) = \vec{SA}^{(1)}_{p}$, which maps the $L_{2}$ MeSH counts to their $L_{1}$ branches.

In order to count co-occurrences, we  take each $\vec{SA}$ and denote the binary SA vector $B_{p} = {\tt Sign}(\vec{SA})$, which reduces each element value to either 0 or 1. For an article with $M_{p}$ non-zero elements, we then count all ${M_{p}\choose{2}}$   pairwise permutations which we record in a normalized co-occurrence matrix, given by  ${\bf M}^{(1)}_{p}$ (${\bf M}^{(2)}_{p}$) for the  $L_{1}$ ($L_{2}$) representation. Each matrix is normalized to unity such that the total of all subelements $\sum_{i,j} {\bf M}_{p,ij} = 1$. This normalization step ensures  that each publication contributes an equal share to the annual co-occurrence matrix, given by ${\bf M}^{(1)}_{y} = \sum_{p \in y} {\bf M}^{(1)}_{p} = N(y)$, where $N(y)$ is the total number of articles analyzed from year $y$ (with similar definition for the $L_{2}$ level).\\

%\noindent{\bf Method for quantifying cluster composition fluctuations.} 

\noindent{\bf Calculation of the cross-cluster bridge score.} 
%{\color{magenta}
Motivated by the bridge centrality index \cite{jensen2016detecting}, we define the knowledge bridge score  $\beta_{i}$ of node $i$, given by
\begin{equation}
\beta_{i}=\sum_{C_{I},C_{J} \in C \ {\tt and} \ C_{I} \neq C_{J}}D_{IJ}W_{iJ} \ ,
\label{BS}
\end{equation}
which first  requires  identifying  nodes belonging to distinct clusters, calculated here using the Louvain algorithm \cite{Blondel:2008}.
Here $C$ represents the set of connected clusters in the giant component of  an undirected network, and $C_{J}$ represents a cluster that is  different from the cluster $C_{I}$ containing a given node $i$. We define $D_{IJ}$ to be the distance between cluster $C_{I}$ and $C_{J}$,  measured as the inverse of the sum of weights of edges between the clusters, represented as $D_{IJ} = W_{IJ}^{-1} = ( \sum_{i \in C_{I} {\tt and} j \in C_{J}} W_{ij})^{-1}$. Similarly, $W_{iJ}$ is the sum of weights of edges between node $i$ and all other nodes belonging to cluster $C_{J}$. For our purposes, we define the link weights as empirical co-occurrence values, $W \equiv {\bf M}^{(2)}_{t}$. Hence,  individual MeSH terms (nodes) that play key roles in bridging knowledge clusters  have high $\beta_{i}$ values; contrariwise,  nodes that are only connected to nodes belonging to their own cluster have $\beta=0$.

Since ${\bf M}^{(2)}_{t}$ tallies are proportional to the total number $N(t)$ of articles published in year $t$, which are generally increasing,   we instead focus on the rank associated with each $\beta_{i}$ score, denoted by $r_{\beta,i}$. {\bf Figure \ref{FigureSIBranchScore.fig}} shows the rank dynamics of the MeSH in each cluster exhibiting significant growth or decline over the study period. To address how to compare ranks of nodes from  clusters of varying size (measured as the  total number of nodes within the cluster $I$, denoted by $\vert C_{I} \vert$), then we  define the normalized Bridge rank score as $R_{i} = r_{\beta,i}/\vert C_{I} \vert$, as plotted in  {\bf Figure \ref{Figure3.fig}}.

Based upon each time series $R_{i,t}$ calculated at the 1-year time resolution, we identified emerging bridges using the following criteria: (i) The node is on average ranked in the top-20 of its own cluster; (ii) the time series is at least half as long as the entire observation period from 1970 to 2018; (ii) There is a significant positive or negative trend, as determined by  linear regression, such that the trend coefficient P-value is smaller than 0.01; (iv) the trend coefficient magnitude is sufficiently large, in magnitude $>$ 0.1. % $r<-0.1$, $-0.01<r<0.01$, or $0.1<r$.
These criteria  identify  8 emerging knowledge bridges: E05, F01, F02, G07, J01, L01, N04 and N05.\\
%}

}%
%\bigskip

 \noindent{\bf Data Availability}: All data analyzed here are openly available from PubMed API, the MeSH tree prescribed by the NLM (see https://meshb.nlm.nih.gov/treeView) and the 2018 Web of Science JCR list of Impact Factors.
%Supporting article-level and  individual-level data will be made available at the Open Science Framework data repository. % \cite{DataOSF}. %https://doi.org/XXX

 \vspace{-0.2in}
\section*{Acknowledgments}
 \vspace{-0.2in}
\noindent  IP acknowledges funding from the Eckhard- Pfeiffer Distinguished Professorship Fund; IP and AMP  acknowledge support from NSF grant 1738163 entitled `From Genomics to Brain Science'. AMP acknowledges financial support from a Hellman Fellow award that was critical to completing this project.  Any opinions, findings, and conclusions or recommendations expressed in this paper are those of the authors and do not necessarily reflect the views of the funding agencies. 

 \vspace{-0.3in}
\section*{Author Contributions}
 \vspace{-0.2in}
\noindent DY  analyzed and visualized the data;  IP  performed  research and participated in the writing of the manuscript; AMP performed the research, drafted the manuscript, collected, analyzed, and visualized the data. 

 \vspace{-0.3in}
\section*{Additional Information}
 \vspace{-0.2in}
\noindent {\bf Supplementary Information} accompanies this paper.

\noindent {\bf Competing financial interests} The authors declare no competing financial interests.
 \vspace{-0.2in}

\bibliography{YPP2020}
\bibliographystyle{naturemag}

\begin{comment}

\end{comment}

\newpage
\clearpage

\newpage
\clearpage

\beginsupplement % TURNS EVERYTHING AFTER HERE INTO SX = Supplemental

%\begin{comment}
\begin{center}
{\bf  \large Supplementary Information -- Appendix Text, Figure S1-S6  \& Table S1} % \& Tables S1-SX
\end{center}
\bigskip
\bigskip
\bigskip

\noindent {\Large \bf Biomedical Convergence Facilitated by  the Emergence of  Technological and Informatic Capabilities} 

\bigskip

\noindent {\bf \large  Dong Yang$^{1}$,   Ioannis Pavlidis$^{2}$, Alexander M. Petersen$^{1}$} 

\bigskip

\noindent $^{1}$Department of Management of Complex Systems, Ernest and Julio Gallo Management Program, School of Engineering, University of California, Merced, California 95343, USA;\\ $^{2}$Computational Physiology Laboratory, University of Houston, Houston, Texas 77204, USA \\

\bigskip\bigskip\bigskip

%\vspace{ -120 pt}
%\noindent {\Large \bf  Appendix Text} 

%\newpage

\noindent {\bf Appendix Text}\\

\section{MeSH cluster dynamics indicate periods of knowledge network reorganization.}
%\noindent{\bf MeSH cluster dynamics indicate periods of knowledge network reorganization.}  
The  knowledge network dynamics, in particular bursts of  cluster size variation exhibited in {\bf Figure \ref{Figure2.fig}}(D),  allude to paradigm shifts mediated by scientific discovery. To identify particularly disruptive periods in the knowledge network,   we developed a network method for identifying fluctuations of individual MeSH constituents across clusters by tracking the entry and exit of constituents between  sequential 1-year periods.  

%In order to be generalizable, this method must account for the fact that clusters themselves are dynamic, and so the largest cluster in one period may have little correspondence with the largest cluster in the following period.
To systematically track relations between MeSH branches, at either the $L_{1}$ or $L_{2}$ level, this straightforward method requires a flexible  system for labeling clusters. 
To this end  we designed a cluster-tagging method that first relies on identifying the stable subclusters, or cliques. In the present case, these cliques correspond to MeSH  that are always observed  in the same cluster together -- in particular within co-occurrence clusters calculated at the 1-year level. 
%See the {\it Methods} section {\it Method for quantifying cluster composition fluctuations}  for further details. 

%%% START METHODS BLIP
%\noindent{\bf Method for quantifying cluster composition fluctuations.} 
{\bf Figure \ref{FigureSIStableCliques.fig}}(A) shows the MeSH cliques for the $L_{1}$ and $L_{2}$ levels. For example, an interesting cross-SA clique observed at the $L_{2}$ level is the dyad formed by ``Technology, Industry, and Agriculture'' (J01) and ``Biomedical and Dental Materials'' (D25); another mono-SA clique is the one formed by seven  $L_{2}$ branches (D01, D02, D03, D04, D09, D10, D27), all from the $L_{1}$ D-branch.  Hence, these stationary cliques are  motifs that uniquely identify an arbitrary cluster $C_{x}$ according to its constituent cliques. 

By way of example, {\bf Fig. \ref{FigureSIStableCliques.fig}}(B) shows the cluster $C_{c}$ which contains the D-clique described above in two subsequent years. Consequently, an arbitrary MeSH denoted by $i$ belonging to $C_{c}$ in year $t$ and $t+1$ is stationary with respect to the cliques. Alternatively, consider an arbitrary MeSH $j$ which was a member of $C_{a}$ (identified by two particular cliques) and $C_{e}$ in the next year (identified by two  cliques, one from $C_{a}$ and one altogether new one). In this case, the MeSH $j$ (indicated by orange)  has partially switched clusters. Finally, consider  an arbitrary MeSH $k$ (indicated by red) which  was a member of $C_{b}$ (identified by a single clique) and $C_{d}$ in the next year (identified by a completely different clique); this case corresponds to minimal continuity (maximal discontinuity) with respect to cluster member cliques. 

To quantify aggregate cluster dynamics, we assign each MeSH (represented generically  by the index $m$) a set of labels $Q_{m,t}= \{q_{1},...,q_{n} \}$ corresponding to each of the $n$ unique cliques present, as indicated by $q$, for a given year $t$. We then calculate the Jaccard distance $\Delta J_{m} = 1- \vert Q_{m,t} \cap Q_{m,t+1} \vert / \vert Q_{m,t} \cup Q_{m,t+1} \vert$. Hence, $\Delta J_{m} =0$ corresponds to maximal continuity (since $\vert Q_{m,t} \cap Q_{m,t+1} \vert  = \vert Q_{m,t} \cup Q_{m,t+1} \vert$). Conversely, $\Delta J_{m} =1$ corresponds to maximal discontinuity (since $\vert Q_{m,t} \cap Q_{m,t+1} \vert  = 0$). When there is a partial shift in cluster member cliques, then we obtain intermediate values, $0 < \Delta J_{m} < 1$.

%{\bf Figure \ref{FigureSIStableCliques.fig}}(C) shows the frequency of $\Delta J_{m}$ values by type at the $L_{1}$ and $L_{2}$ levels, which are consistent in terms of the periods in which there are the highest levels of MeSH network reorganization, characterized by $\Delta J_{m} > 0$ values (indicated by orange and red curves). \\
%%% END METHODS BLIP

{\bf Figure \ref{FigureSIStableCliques.fig}}(C) indicates the periods with the highest levels of MeSH network reorganization, characterized by $\Delta J_{m} > 0$ values (indicated by orange and red curves). 
These results are consistent for both the e $L_{1}$ and $L_{2}$ resolution levels.
 MeSH data aggregated at the $L_{1}$ level indicates a period of heightened MeSH cluster dynamics starting in the roughly 2000 and continuing in bursts up to present. Additional inspection at the $L_{2}$ level indicates periodic fluctuations occurring over time, but again with heightened turbulence  in the early 1980s, and in the years around 2010. We interpret these heightened periods of MeSH cluster discontinuity as indicators of individual MeSH that emerge as cross-cluster knowledge bridges that integrate distinct knowledge domains, a hallmark of biomedical convergence in topical SA space. As such, the objective of this method is to the objective of which is to identify and justify disruptive periods analyzed in detail in the following sections. \\
 
 \section{Techno-informatic capabilities facilitate biomedical convergence around brain, behavior and health science.}
 %\noindent{\bf Techno-informatic capabilities facilitate biomedical convergence around brain, behavior and health science.} 
Concurrent co-occurrence of MeSH pairs (denoted by $j$ and $j'$) that are close neighbors of a given MeSH (i.e., the ego node, $i$) is indicative of recombinant knowledge domains  mediated by $i$. To analyze this triadic closure phenomena as it relates to cross-domain convergence, we explored dynamic patterns occurring in the subset of  ${\bf M}^{(2)}_{jj'}$  values among the most prominent co-occurring neighbors of a given MeSH. 
In particular, we focused on the  co-occurrence dynamics  for each of the  convergence bridges identified  in {\bf Fig. \ref{Figure3.fig}}. 

For each convergence bridge MeSH, we then selected the ten most frequently co-occurring MeSH for the denoted period.
{\bf Figure \ref{FigureSINNetworks.fig}} shows the subset of ${\bf M}^{(2)}$ values visualized as a  subnetwork. 
For a given period, the top ten co-occurring neighbors ($j$) are sorted in clockwise fashion starting from the top, with nodes sized proportional to  ${\bf M}^{(2)}_{ij}$. To facilitate visual inspection,  each MeSH's SA  is indicated by  its node/label color.
Links are plotted with thickness and shade proportional to ${\bf M}^{(2)}_{jj'}$, thereby indicating the cross-domain linkages among prominent neighbors that are facilitated by $i$.   Each node includes its MeSH identifier and a knowledge cluster identifier, the latter indicated by a gray  scale gradient. 

The most significant feature of each convergence bridge are as follows:   ``Investigative Techniques'' (E05) integrate all clusters with high SA diversity.  ``Behavior and Behavior Mechanisms'' (F01)
integrates health, ``Physiological Phenomena'', ``Eukaryotes'' and ``Pathological Conditions, Signs and Symptoms'' domains. 
%[F02] ``Psychological Phenomena'' is converging with basic research on ``Musculoskeletal and Neural Physiological Phenomena''. 
 F01 and  ``Psychological Phenomena'' (F02)  increasingly incorporate ``Information Science'' (L01) and ``Investigative Techniques''.
 ``Physiological Phenomena'' (G07),  which exhibits a high co-occurrence with ``Food and Beverages'' research, representing the increasingly lucrative domain of human health science, also exhibits high diversity of cross-SA and cross-cluster.
 ``Technology, Industry, and Agriculture''  (J01) increasingly integrates from the 1990s onward with the ``Chemical and Drugs'' domain, highlighting the role of academic-industry-government   cross-industry triple-helix \cite{etzkowitz2000dynamics,leydesdorff1996emergence} underlying the pharmaceutical industry.
L01 integrates with ``Mathematical Concepts'' to facilitate research investigating ``Behavior and Behavior Mechanisms'' by leveraging technology providing non-invasive ``Investigative Techniques'.
And both  ``Health Services Administration'' (N04) and  ``Health Care Quality, Access, and Evaluation'' (N05), which have become highly coupled mirrors of each other, have in the last decade incorporated ``Information Science'' methods for clinical diagnosis and classification of ``Pathological Conditions, Signs and Symptoms'' and overall health care program assessment. 

\section{Outer-product method for measuring  categorical  diversity and disparity. } 
 We leverage a generic tensor-product method that takes as input a weighted vector and yields a scalar diversity measure based upon categorical mixing.  
  While the result of our approach is nearly identical to the Blau index (also referred to as the Gini-Simpson index), our diversity measure is motivated by way of dyadic co-occurrence rather than the standard formulation  motivated around repeated  sampling.  
 Within the present context, the weighted vector of category counts is the SA decomposition of an article, given by $\vec{SA}_{p}^{(1)}$ or $\vec{SA}_{p}^{(2)}$.  The resulting metric, represented by $f_{D,p}$, quantifies the degree of  cross-domain co-occurrence and is a bounded statistic in the range $[0,1)$.

Calculating $f_{D,p}$ begins with the outer-product $\vec{SA}_{p} \otimes  \vec{SA}_{p}$, where $\otimes$ is the outer tensor product. The resulting matrix represents dyadic combinations of categories as opposed to permutations (i.e., capturing the subtle difference between an  undirected  and directed network). While we did not explore it further, this matrix formulation may also give rise to higher-order measures of diversity associated with the eigenvalues of the outer-product matrix. 

We   normalize the resulting outer-product matrix so that our analysis is not systematically biased by increase  in the number of MeSH per paper over time; see \cite{petersen2016triple} for secular growth in the number of MeSH per article, reflecting growth of the MeSH ontology accompanied by growth in the length and breadth of published research. 
All steps together, the normalized co-occurrence matrix  ${\bf D}_{p}$ is given by
\begin{eqnarray}
{\bf D}_{p} \equiv \frac{U(\vec{SA}_{p} \otimes  \vec{SA}_{p})}{\vert\vert U(\vec{SA}_{p} \otimes  \vec{SA}_{p}) \vert\vert} \ . 
\label{Dmatrix}
\end{eqnarray}
Without loss of generality,  this definition involves $U({\bf G})$, an operator yielding the upper-diagonal elements of the arbitrary matrix ${\bf G}$ (i.e., representing the undirected co-occurrence network among  subcategories);   $\vert\vert ... \vert\vert$ indicates the matrix total calculated by summing across all matrix elements. 
The subtle difference between the Blau index arises from $U({\bf G})$, which is imposed to capture the difference between combinations rather than permutations  (or directed versus undirected network). Hence, this perspective offers a new pathway to this fundamental diversity measure by way of co-occurrence rather than repeated sampling.

Hence, ${\bf D}_{p}(\vec{SA}_{p})$ tabulates the weighted product across  all undirected SA-SA pairs.
We then define the co-occurrence   diversity  as the scalar quantity given by 
\begin{eqnarray}
f_{D,p}=1-\mathrm{Tr}({\bf D}_{p}) \in [0,1) \ ,
\label{fDp}
\end{eqnarray}
where $\mathrm{Tr}$ is the matrix trace, corresponding to  the sum of  diagonal elements in ${\bf D}_{p}$. Hence, $f_{D,p}$ is the tally of off-diagonal elements in ${\bf D}$.  
 Articles featuring a single SA have value $f_{D,p} = 0$, whereas articles featuring multiple SA have values in the range $0 < f_{D,p} < 1$. Regarding the upper limit,  when all vector elements have  equal values then $f_{D,p}=(d-1)/(d+1)$, where  $d$ is the dimension of the categorical vector (for the Blau index the upper limit is instead d-1/d).  In the present case $d=10$ and so the maximum $f_{D,p}$ value is 9/11. 
The average article diversity by publication year,  denoted by $\langle f_{D}(t) \rangle$, is representative of a characteristic article since $f_{D,p} \in [0,1)$ is  bounded. 
 
By way of example, consider a decomposition across only 6 SA, and the particular case of an article with 4 MeSH   belonging to 3 different SA, e.g.  $\vec{SA}_{p}=\{1,2,0,0,1,0\}$. Calculation of  the co-occurrence matrix ${\bf D}_{p}(\vec{SA}_{p}) $  in Eq. (\ref{Dmatrix}) yields    

\bigskip
${\bf D}_{p}(\vec{SA}_{p}) =$ 
\(\left(
\begin{array}{cccccc}
 \frac{1}{11} & \frac{2}{11} & 0 & 0 & \frac{1}{11} & 0 \\
 \\
 0 & \frac{4}{11} & 0 & 0 & \frac{2}{11} & 0 \\
  \\
 0 & 0 & 0 & 0 & 0 & 0 \\
  \\
 0 & 0 & 0 & 0 & 0 & 0 \\
  \\
 0 & 0 & 0 & 0 & \frac{1}{11} & 0 \\
  \\
 0 & 0 & 0 & 0 & 0 & 0 \\
\end{array}
\right)\)
\bigskip

\noindent with $\vert\vert U(\vec{SA}_{p} \otimes  \vec{SA}) \vert\vert=11$. The categorical  diversity is the  total across the off-diagonal elements,  $f_{D,p}= 5/11$.

For comparison, consider the representation of an article with the same number of metadata entities that all fall into just the second category,  $\vec{SA}_{p}=\{0,4,0,0,0,0\}$. In which case

\bigskip
${\bf D}_{p}(\vec{SA}_{p}) =$ 
\(\left(
\begin{array}{cccccc}
 0 & 0 & 0 & 0 &0 & 0 \\
 \\
 0 &1 & 0 & 0 & 0 & 0 \\
  \\
 0 & 0 & 0 & 0 & 0 & 0 \\
  \\
 0 & 0 & 0 & 0 & 0 & 0 \\
  \\
 0 & 0 & 0 & 0 & 0& 0 \\
  \\
 0 & 0 & 0 & 0 & 0 & 0 \\
\end{array}
\right)\)
\bigskip

\noindent  and so $f_{D,p} = 1 - 1 = 0$. 

 What does this measure measure? Notably, $f_{D,p}$ accounts for both  categorical differences (Shannon-like) and concentration disparity (Gini-like)  \cite{harrison2007s}.  One the first hand, articles with more variation in SA categories will correspond to larger $f_{D,p}$ values, as the number of non-zero off-diagonal elements is proportional to ${{M_{p}}\choose{2}}\sim M_{p}^{2}$, where $M_{p}$ is the number of distinct SA present, which contributes to larger $f_{D,p}$; and on the second hand, the off-diagonal elements will be relatively larger in combination if the count values contained in $SA_{2}$ are more evenly distributed, i.e., are not  highly concentrated in just one category.

%\newpage
%\clearpage

\begin{figure*}[b!]
\centering{\includegraphics[width=0.89\textwidth, angle = 0]{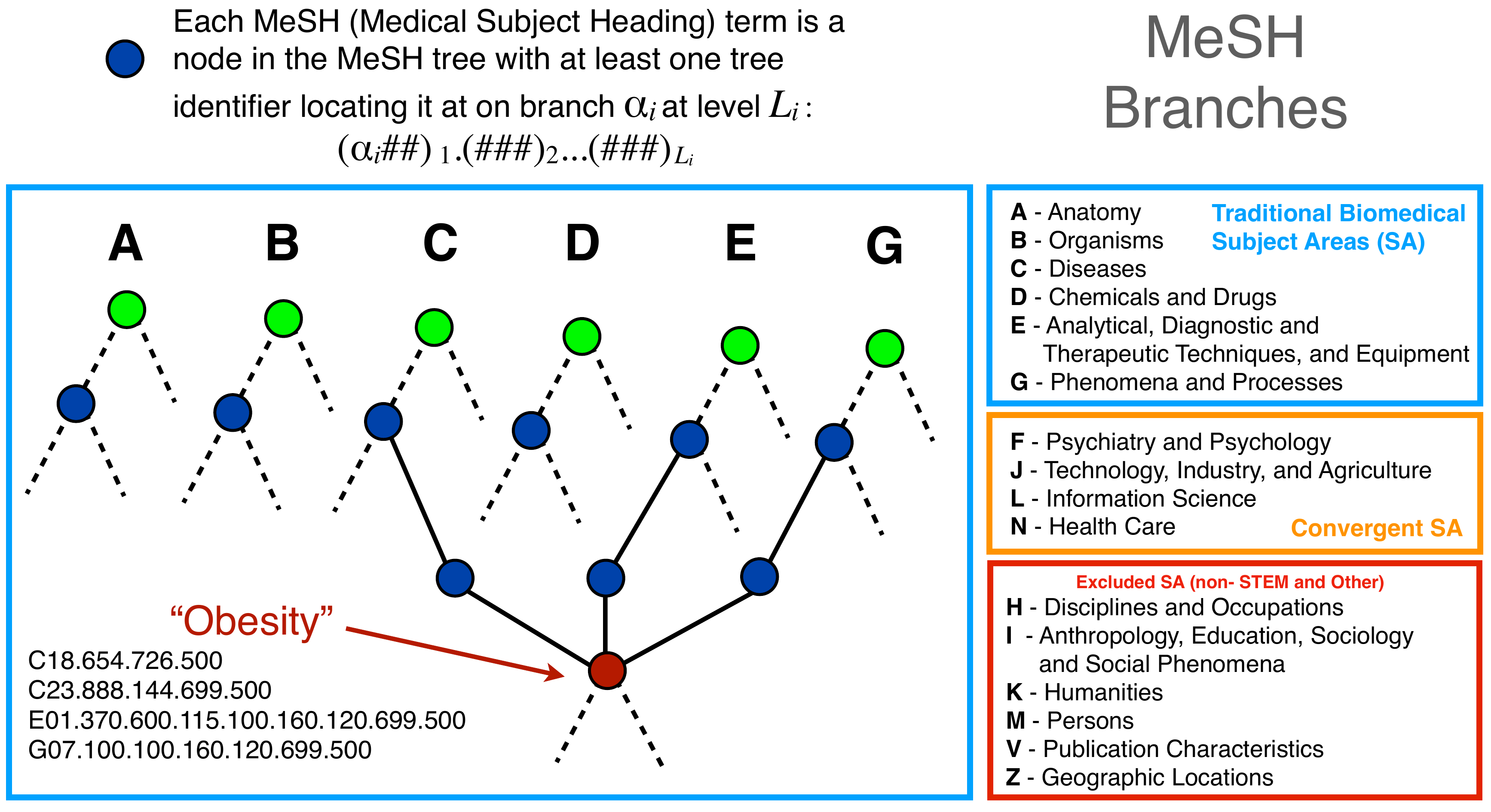}}
 \caption{{\bf Schematic representation of the quasi-hierarchical MeSH tree.} 
 We restrict this illustration to MeSH belonging to the core biomedical branches $ \alpha =\{A,B,C,D,E,G\}$. 
 The location of each MeSH $m$ is given by the set of locations of the parent nodes directly above it. For example the MeSH ``Overnutrition'' with Tree Number C18.654.726 is located at the third level, and has daughter MeSH ``Obesity''  which has four Tree Numbers corresponding to three distinct branches (C, E and G): C18.654.726.500, C23.888.144.699.500, E01.370.600.115.100.160.120.699.500, G07.100.100.160.120.699.500. Hence, this particular MeSH maps onto three $L_{1}$ branches represented by $\vec{SA}^{(1)}=$ \{0,0,2,0,1,0,1,0,0,0\}. 
 } 
 \label{Mtree.fig}
\end{figure*} 

%\begin{figure*}
%\centering{\includegraphics[width=0.79\textwidth, angle = 0]{Method_Fig3.pdf}}
% \caption{The MeSH tree is not a true tree: (i) loop degeneracies and (ii) non-uniform levels. } 
% \label{Mloops}
%\end{figure*}  

%\begin{figure*}
%\centering{\includegraphics[width=0.99\textwidth, angle = 0]{Mesh_Branch_A.pdf}}
% \caption{Network visualization of branch A (``Anatomy'') of the  MeSH classification tree. } 
% \label{MeshBranchA}
%\end{figure*}  

\begin{figure*}
\centering{\includegraphics[width=0.99\textwidth, angle = 0]{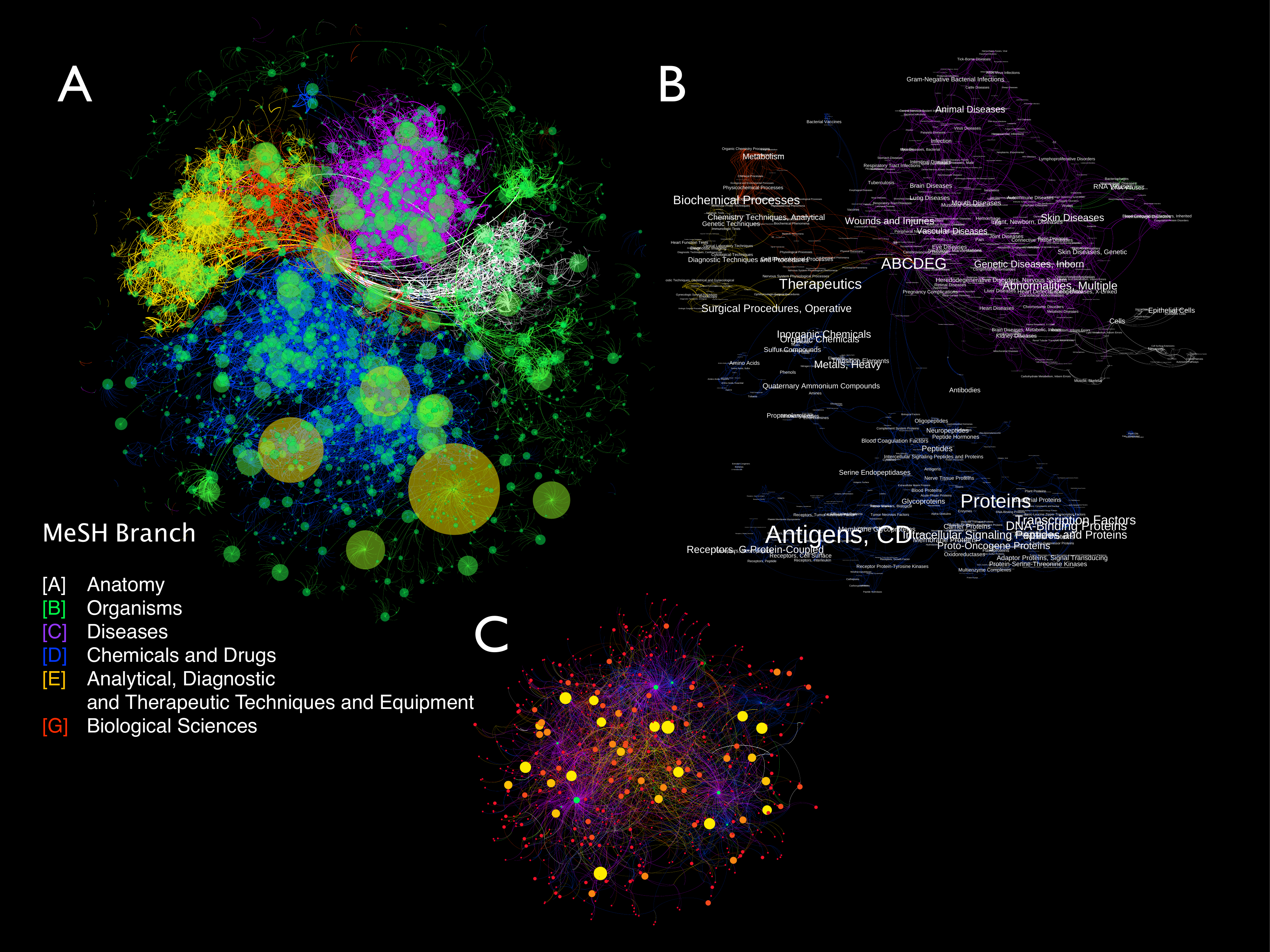}}
 \caption{{\bf Structure and interconnectivity of the  MeSH knowledge network.}(A)  Network visualization of the MeSH branches $\alpha_{i} = \{ A, B, C, D, E, G\}$. Size and color  of nodes is proportional to the in-degree  of the MeSH term (from higher tree levels).  The color of the link is associated with the branch classification of the parent node. (B) MeSH network with most prominent nodes labeled for visual inspection. (C) The  network of links that span different tree branches, $\alpha_{i} \leftrightarrow \alpha_{j}$. } 
 \label{MeshBranchABCDEG.fig}
\end{figure*}  

\begin{figure*}[!t]
\centering{\includegraphics[width=0.99\textwidth]{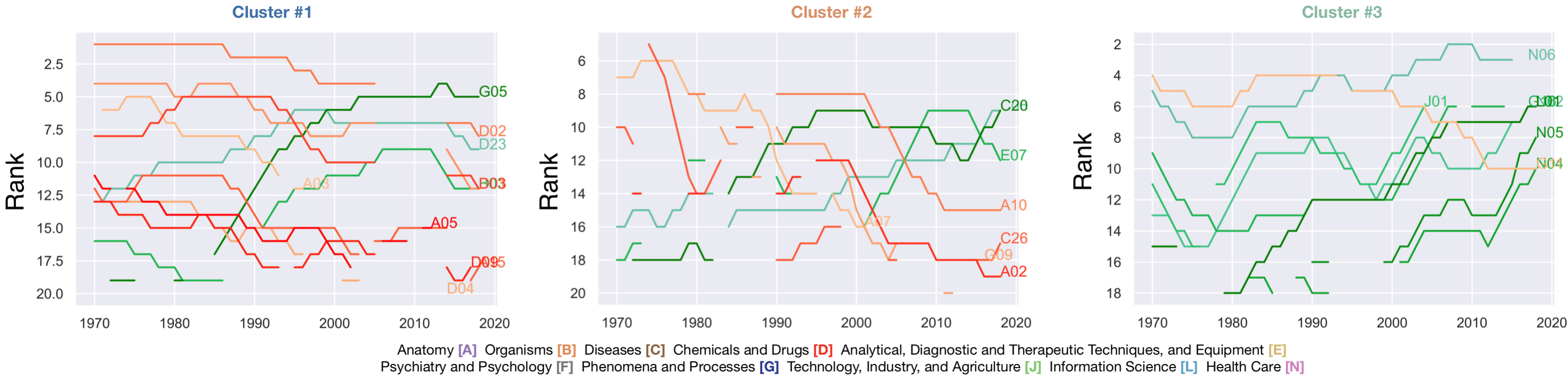}}
 \caption{  \label{FigureSIBranchScore.fig} {\bf MeSH with high bridge rank and significant trends -- by knowledge cluster.} 
 Evolution of knowledge bridge ranks for individual MeSH that feature either sustained high bridge rank  or significant positive (indicated by green) or negative (indicated by red) rank change over the entire period of analysis.  
 }
\end{figure*} 

\begin{SCfigure}
\includegraphics[width=0.65\textwidth]{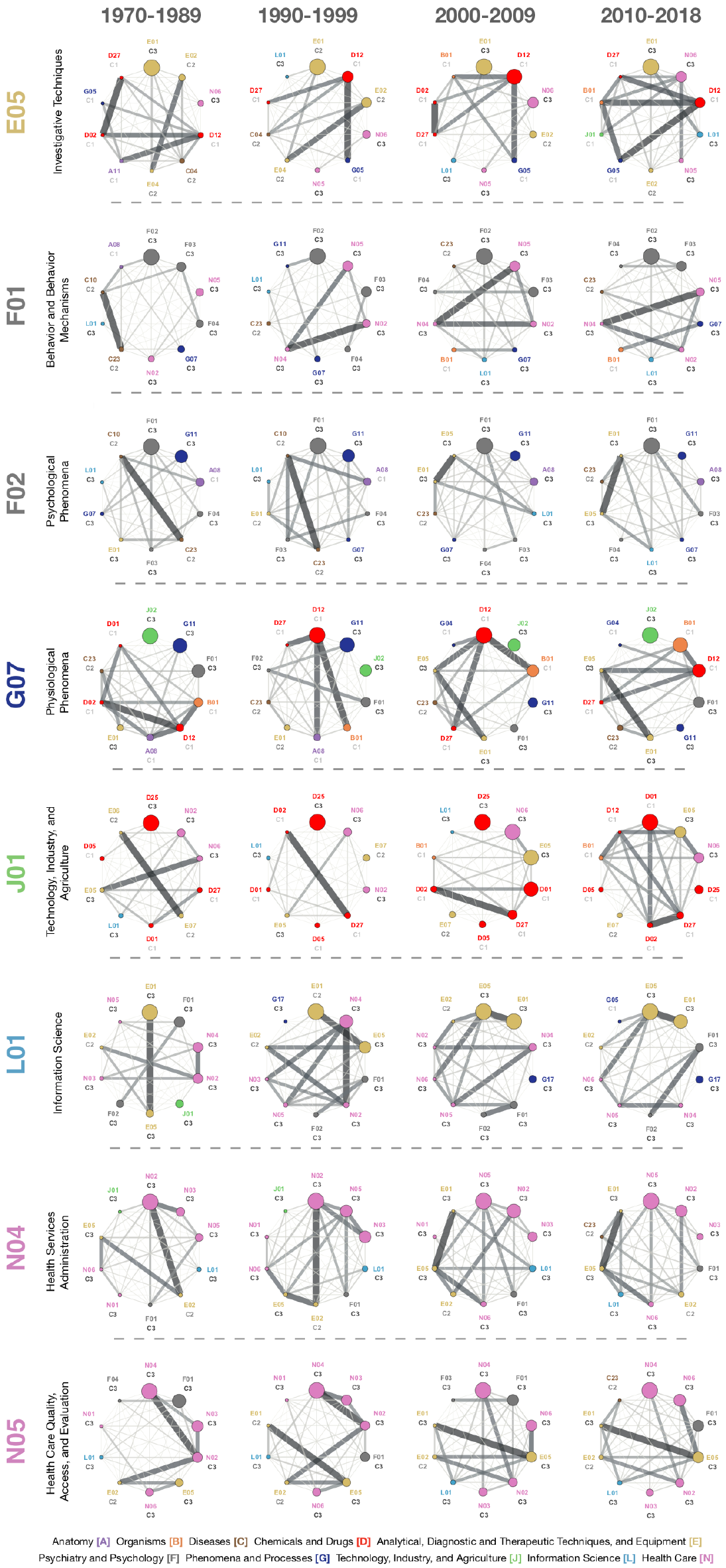}
\caption{ \label{FigureSINNetworks.fig}    {\bf Temporal network dynamics for  8 convergence bridges.} 
Each row illustrates a MeSH term ($i$) identified as a convergence bridge. Each network is calculated using data from the indicated period,  and shows the ten most frequently co-occurring MeSH  -- sorted clockwise, starting from the top, with nodes sized proportional to  ${\bf M}^{(2)}_{ij}$; each MeSH's SA  is indicated by  its node/label color.
Links are plotted with thickness and shade proportional to ${\bf M}^{(2)}_{jj'}$, thereby indicating the cross-domain linkages among prominent neighbors that are facilitated by $i$.   Each node includes its MeSH identifier and a knowledge cluster identifier, the latter indicated by a gray  scale gradient. 
For example, MeSH J01 ``Technology, Industry, and Agriculture'' (which is a member of $C3$ for the first three periods and subsequently transitioning to $C1$ in the most recent period) is highly connected to MeSH from all other  clusters (C1-C3), in particular to L01 until its disassociation in the most recent period 2010-2018; interestingly, L01 ``Information Science'' diverged from $J01$ as early as the first period 1970-1989, subsequently becoming more strongly coupled with members of branch E and G.   }
\end{SCfigure}

\begin{figure*}[!t]
\centering{\includegraphics[width=0.99\textwidth]{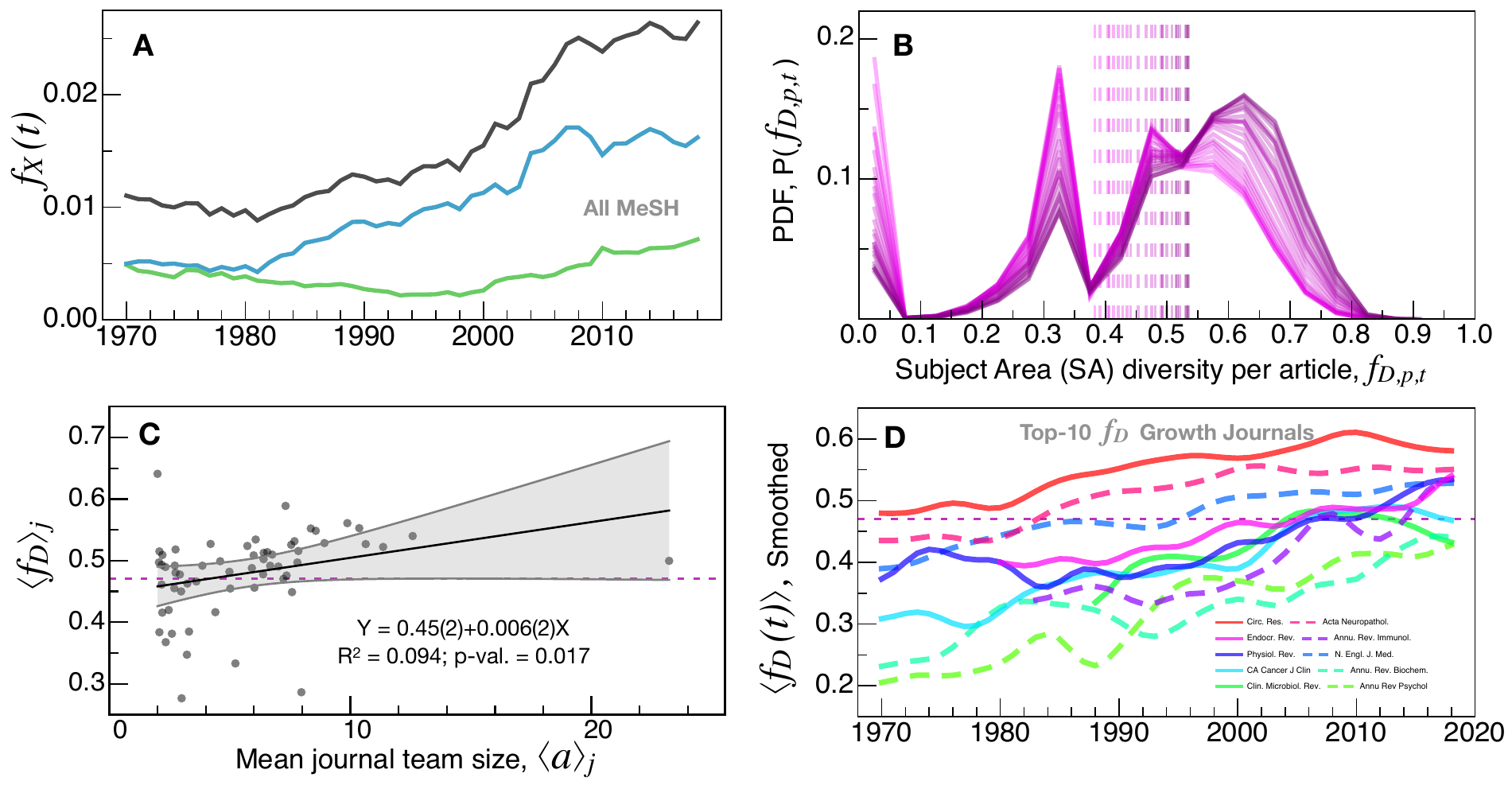}}
 \caption{   \label{FigureSIfD.fig} {\bf   Cross-domain  knowledge integration -- supporting analysis.} 
(A)  Analog to Fig. \ref{Figure4.fig}(A) but calculated using all MeSH (i.e., analyzing MeSH independent of Major/Minor status).
(B) Probability distribution $P(f_{D,p})$ for each of the 3-year subsamples analyzed in  Fig. \ref{Figure4.fig}(B). One notable shift is the decreased prevalence of mono-domain articles ($f_{D,p}=0$) which is a large but not sole contributor to the increasing trend in average value, $\langle f_{D} (t) \rangle$ (shown as horizontal vertical dashed lines). Nevertheless, the right tail of the distribution is well defined and does not increase dramatically in range (i.e., the max diversity value is persistently  around the value 0.9). The peak around value 1/3 (1/2) corresponds to articles with  MeSH contributing equally to two (three) distinct SA and represents the first (second) mode of cross-domain convergence. 
(C)  Robustness check for Fig. \ref{Figure4.fig}(C), showing positive relationship between average  team size and cross-domain SA diversity calculated at the journal level  (ANOVA p-val.= 0.017; shaded region is 99\% CI).
(D) Ten  biomedical journals featuring the highest growth in average $f_{D,p}$ values over 1970-2018.
}
\end{figure*} 

\begin{SCfigure}
\includegraphics[width=0.65\textwidth]{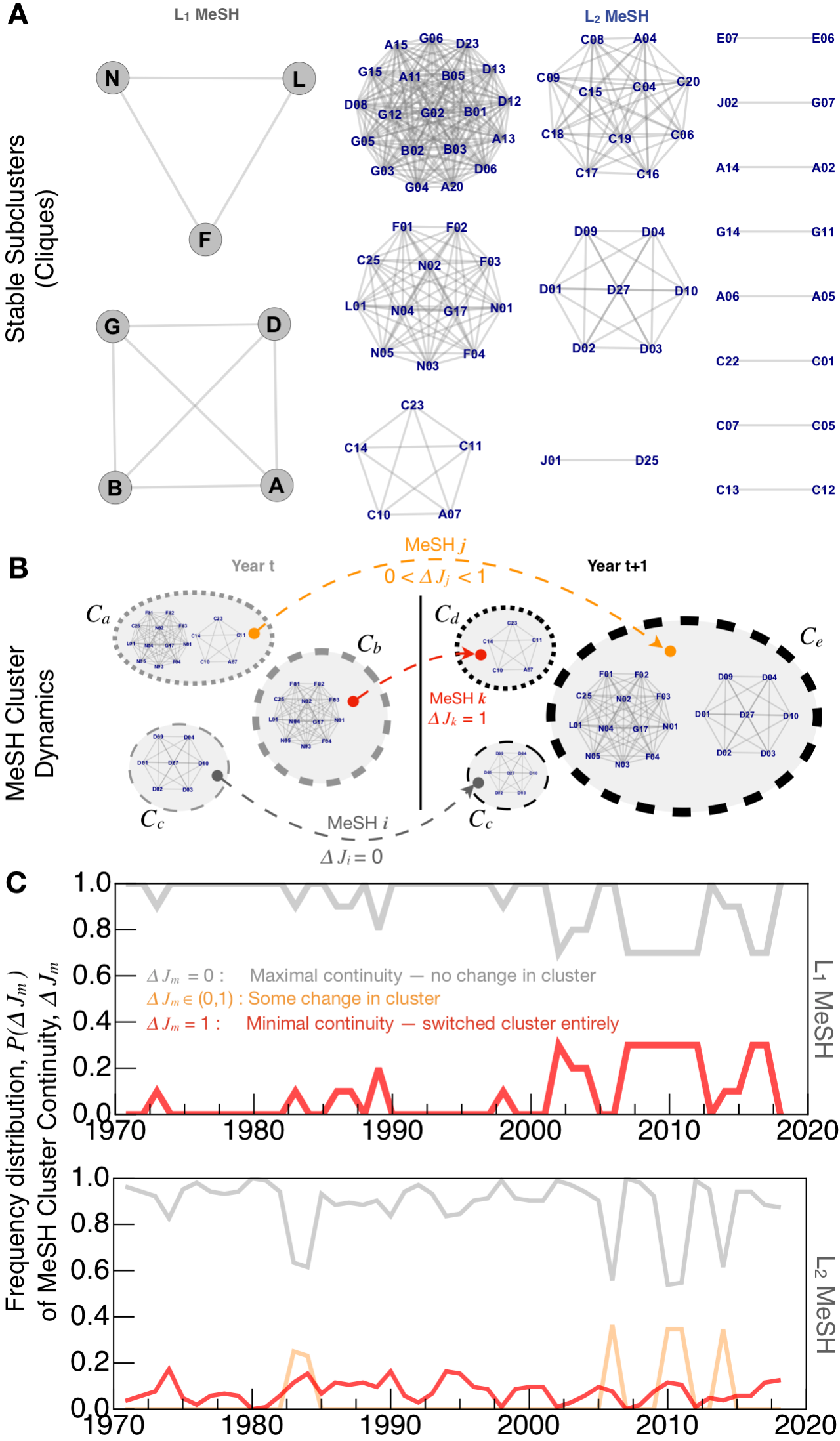}
\caption{ \label{FigureSIStableCliques.fig}    {\bf Method for quantifying fluctuations of individual MeSH across knowledge clusters.} Reorganization of co-occurrence networks are indicative of paradigmatic disruption and cross-domain bridge formation. Stable subclusters are groups of MeSH that always appear together in annual-level co-occurrence matrix Louvain clustering -- see {\bf Fig.} \ref{Figure2.fig}(C). These cliques are used as an unique identification system for calculating the cluster continuity $\Delta J_{m}$ of a given MeSH term $m$ over sequential 1-year periods. (A) Cliques for the period 1970-2018 calculated for  the $L_{1}$ (left) and $L_{2}$ (right) levels. (B) Schematic of the identification system for tracking cluster dynamics of individual $m$. For example, comparing MeSH $i$ for the sequential year $t$ and $t+1$ periods, $i$ stayed in the cluster $C_{c}$ (identified as having the D-clique comprised of seven $L_{2}$ MeSH from the D-branch)  and so there is maximal continuity of the cliques defining its surrounding cluster. Contrariwise, MeSH $k$ transitioned from a cluster in $t$ defined by a single clique that differs from the clique defining its cluster in $t+1$; hence, this case corresponds to minimal continuity. The case of MeSH $j$ is inbetween these extreme cases, whereby $j$ transitions from a cluster in $t$ that shares a common clique as the cluster in $t+1$, but with differing second member clique. (C) Frequency distribution of continuity values $\Delta J_{m}$ by year. Years with low MeSH cluster continuity feature higher frequencies of MeSH that switch knowledge clusters entirely (red) or partially (orange); conversely, periods with high cluster continuity, in which all $m$ stay in the same cluster, corresponds to the gray curve approaching unity. The period with the highest levels of  knowledge cluster discontinuity started in the early-2000s and peaked around 2010. Discontinuity peaks  indicate the emergence of  individual MeSH serving as cross-cluster bridges. }
\end{SCfigure}

\begin{table}[b!]
\caption{ {\bf Top-20  Convergent Journals.} Biomedical journals in the  top-100 2018 JCR Impact Factor, ranked  according to average categorical SA diversity  over the period 1970 to 2018. }
%\small
\resizebox{0.29\columnwidth}{!}{
\begin{tabular}{@{\vrule height 10.5pt depth4pt  width0pt}l|c|c|}
&\multicolumn1c{ % OLS parameter estimates
}\\
\hline
\noalign{
\vskip-0pt}
\vrule depth 6pt width 0pt Journal ($j$)    & Average SA diversity   \\
\vrule depth 6pt width 0pt    & $\langle f_{D} \rangle_{j,1970-2018}$  \\
\hline  
\hline  
Circ. Res.   &   0.556\\
J. Clin. Oncol.   &   0.549\\
J. Am. Coll. Cardiol.   &   0.511\\
Acta. Neuropathol.   &   0.509\\
Nat. Genet.   &   0.506\\
Gut   &   0.493\\
PNAS   &   0.483\\
Science   &   0.481\\
Ann. Intern. Med.   &   0.474\\
N. Engl. J. Med.   &   0.466\\
Cell   &   0.466\\
Nature   &   0.464\\
Eur. Heart J.   &   0.462\\
BMJ   &   0.459\\
Endocr. Rev.   &   0.456\\
Lancet   &   0.452\\
Clin. Microbiol. Rev.   &   0.436\\
Physiol. Rev.   &   0.434\\
Annu. Rev. Immunol.   &   0.400\\
CA. Cancer J. Clin.   &   0.389\\
%Pharmacol. Rev.   &   0.385\\
%Trends Biochem. Sci.   &   0.361\\
%Annu. Rev. Psychol.   &   0.342\\
%Annu. Rev. Biochem.   &   0.327\\
\hline
\hline
\end{tabular}
} %end  \resizebox{1.65\columnwidth}{!}{
\label{table:top20}
\end{table}

\end{document}